\documentclass[11pt,superscriptaddress,amsmath,amssymb,nofootinbib,longbibliography,preprint]{revtex4-1}

\usepackage{graphicx}
\usepackage{dcolumn} 
\usepackage{bm}
\usepackage{amssymb}
\usepackage{amsmath} 
\usepackage{color}
\usepackage[normalem]{ulem}
\usepackage{hyperref}
\usepackage{slashed}
\usepackage{wrapfig}

\newcommand{\Slash}[1]{{\ooalign{\hfil/\hfil\crcr\(#1\)}}}
\allowdisplaybreaks[1]

\begin{document}

\title{Next-to-leading order corrections to decays of the heavier CP-even Higgs boson \\in the two Higgs doublet model}


\preprint{OU-HET 1113, NU-EHET 001}

\author{Shinya Kanemura}
\email{kanemu@het.phys.sci.osakau.ac.jp}
\affiliation{Department of Physics, Osaka University, Toyonaka, Osaka 560-0043, Japan}

\author{Mariko Kikuchi}
\email{kikuchi.mariko13@nihon-u.ac.jp}
\affiliation{College of Engineering, Nihon University, Koriyama, Fukushima 963-8642, Japan}

\author{Kei Yagyu}
\email{yagyu@het.phys.sci.osaka-u.ac.jp}
\affiliation{Department of Physics, Osaka University, Toyonaka, Osaka 560-0043, Japan}

\begin{abstract}
  We investigate the impact of electroweak (EW), scalar and QCD corrections to 
 the full set of decay branching ratios of an additional CP-even Higgs boson ($H$) in the two Higgs doublet model with a softly broken $Z_2$ symmetry.   
 We employ the improved gauge independent on-shell scheme in the renormalized vertices.  
 We particularly focus on the scenario near the alignment limit 
 in which couplings of the discovered 125 GeV Higgs boson ($h$) coincide with those of the standard model 
 while the $Hhh$ coupling vanishes at tree level.
  The renormalized decay rate for $H\to hh$ can significantly be changed from the prediction at tree level
  due to non-decoupling loop effects of additional Higgs bosons, even in the near alignment case. 
 We find that the radiative corrections to the branching ratio of $H\to hh$ can be a few ten percent level in the case with the masses
 of additional Higgs bosons being degenerate  
 under the constraints of perturbative unitarity, vacuum stability and the EW precision data. 
 Further sizable corrections can be obtained for the case with a mass difference among the additional Higgs bosons.
  \end{abstract}

\maketitle


\section{Introduction}\label{sec:Intro}
Current data at LHC show that properties of the discovered Higgs boson ($h^{}$)~\cite{ATLAS:2012yve,CMS:2012qbp} are consistent with those of the standard model (SM) Higgs boson within
experimental uncertainties~\cite{ATLAS:2019nkf,CMS:2020gsy}. 
In addition, there is so far no report which clearly indicates signatures of new particles from collider experiments. 
On the other hand, new physics beyond the SM must exist, because there are phenomena which cannot be explained in the SM such as
neutrino oscillations, dark matter and baryon asymmetry of the Universe. 
In such new physics models, the Higgs sector is often extended from the minimal one assumed in the SM,
and its structure depends on new physics scenarios. 
Therefore, clarifying the structure of the Higgs sector is important to determine the direction of new physics. 

One of the useful ways to determine the structure of extended Higgs sectors
is to measure deviations in various observables of $h^{}$ from SM predictions.  
We can extract information on the structure such as the number of additional Higgs fields and their representations  
from the pattern of the deviations~\cite{Kanemura:2014bqa}. 
Furthermore, the mass scale of extra Higgs bosons can be deduced from the size of the deviations~\cite{Kanemura:2004mg,Kanemura:2014dja,Kanemura:2015mxa,Kanemura:2017wtm,Kanemura:2018yai,Kanemura:2019kjg,Krause:2016oke,Krause:2016xku,Krause:2019qwe,Arhrib:2003vip}. 
Because the precise measurements of cross sections, decay branching ratios (BRs) and the width of the Higgs boson will be performed
at future collider experiments such as the high luminosity LHC (HL-LHC)~\cite{Cepeda:2019klc}, International Linear Collider (ILC)~\cite{Baer:2013cma,Fujii:2017vwa,Asai:2017pwp,LCCPhysicsWorkingGroup:2019fvj}, Circular Electron Positron Collider (CEPC)~\cite{CEPC-SPPCStudyGroup:2015csa} and Future Circular Collider (FCC-ee)~\cite{TLEPDesignStudyWorkingGroup:2013myl}, 
precise predictions of these observables including radiative corrections are necessary to compare these measurements. 
There are several public tools to calculate radiative corrections in models with an extended Higgs sector; i.e.,
\texttt{H-COUP}~\cite{Kanemura:2017gbi,Kanemura:2019slf}, \texttt{2HDECAY}~\cite{Krause:2018wmo} and \texttt{Prophecy4f}~\cite{Denner:2019fcr}. 

Another way is direct searches for additional Higgs bosons.
The current situation mentioned above~\cite{ATLAS:2020zms,ATLAS:2019tpq,ATLAS:2018rvc,ATLAS:2018rnh,ATLAS:2017uhp,ATLAS:2020tlo,ATLAS:2020qiz,ATLAS:2020gxx,ATLAS:2021upq,CMS:2018hir,CMS:2019pzc,CMS:2017ucf,CMS:2017aza,CMS:2019bnu,CMS:2019kca,CMS:2020imj,CMS:2019bfg} would indicate that additional Higgs bosons are too heavy to be directly detected,
or otherwise their masses stay at the electroweak (EW) scale but their decay products are hidden by huge backgrounds. 
The latter possibility suggests that the Higgs sector is nearly aligned without decoupling~\cite{Gunion:2002zf,Craig:2013hca,Carena:2013ooa,Haber:2015pua,Bernon:2015qea}, in which couplings of the discovered Higgs boson are  aligned to those of the SM. 
In this case, decays of additional Higgs bosons into a lighter Higgs boson, ``Higgs to Higgs decays'', 
are suppressed due to the small mixing with $h^{}$, which can be regarded as a golden channel for the direct searches, see e.g., $H\to hh$ and $A\to Zh$ in Refs.~\cite{Dumont:2014wha,Craig:2015jba,Bernon:2015qea,Bernon:2015wef,Chowdhury:2017aav,Su:2019dsf,Kling:2020hmi, Kanemura:2014dea,Aiko:2020ksl}. 
It has been shown in Ref.~\cite{Aiko:2020ksl} that by considering the synergy between direct searches for additional Higgs bosons at the HL-LHC 
and precise measurements of the $h$ couplings at the ILC, wide regions of the parameter space on extended Higgs sectors can be explored.

Such a synergy analysis discussed in Ref.~\cite{Aiko:2020ksl} has been performed at leading order (LO) in EW interactions.  
However, tree level analyses might not be sufficient because of the following reasons.
First, they have analyzed the excluded and testable parameter regions inputting the tree level $hVV$ coupling ($V=W, Z$) as constraints from indirect searches.
If we take into account loop corrections, the $hVV$ coupling is modified. 
Second, decay rates of the Higgs to Higgs decay processes can be significantly changed at one-loop level from the prediction at tree level,  
because some of one-loop diagrams do not vanish at the alignment limit and can be sizable due to the non-decoupling effects of additional Higgs bosons.
Therefore, in order to see the synergy in a more realistic way, 
radiative corrections to both the discovered Higgs boson couplings and the decay BRs of additional Higgs bosons should be taken into account. 
Radiative corrections to decays of additional Higgs bosons have been studied
in Refs.~\cite{Barger:1991ed,Osland:1998hv,Philippov:2006th,Williams:2007dc,Williams:2011bu,Krause:2016xku} for a heavy CP-even Higgs boson,
in Refs.~\cite{Krause:2019qwe,Chankowski:1992es} for a CP-odd Higgs boson and
in Refs.~\cite{Akeroyd:1998uw,Akeroyd:2000xa,Santos:1996hs,Krause:2016oke,Aiko:2021can} for singly charged Higgs bosons.

In this paper, we investigate the impact of next-leading-order (NLO) corrections in EW and scalar interactions to 
the decay BRs of an additional CP-even Higgs boson ($H$) in the two Higgs doublet model (THDM) with a soft broken $Z_2$ symmetry~\cite{Glashow:1976nt, Paschos:1976ay} as a simple but important example. 
We compute decay rates of $H\to hh$, $H\to f\bar{f}$, $H\to VV$, $H\to \gamma\gamma$, $H\to Z\gamma$ and $H\to gg$ at one-loop level
based on the improved on-shell renormalization scheme without gauge dependences, while QCD corrections are also included in decay rates of $H\to q\bar{q}$, $H\to \gamma\gamma$, $H\to Z\gamma$ and $H\to gg$. 
We study the relation between model parameters; e.g., such as masses of additional Higgs bosons and mixing parameters, and the effects of radiative corrections for the $H\to hh$ decay.
\footnote{In Refs.~\cite{Krause:2016xku,Krause:2019qwe}, one-loop corrections to two-body decay rates of $H$ have been calculated, especially focusing on differences among various renormalization schemes. } 
Moreover, we clarify how the one-loop corrections can significantly change the tree level predictions in the scenario with nearly alignment, 
and the correlation between the deviation of the $h\to WW^*$ decay from the SM prediction and the one-loop corrected decay rate of $H\to hh$ under 
theoretical and experimental constraints.
We also study the correlation between the BR of the $H\to hh$ decay and the deviation of the one-loop corrected $hhh$ vertex of the THDM
~\cite{Kanemura:2004mg,Kanemura:2015mxa,Kanemura:2017wtm}
from that of the SM, which is important for testing the EW baryogenesis scenario~\cite{Kanemura:2002vm,Kanemura:2004ch,Grojean:2004xa,Braathen:2019zoh,Braathen:2020vwo}. 
Finally, we give results of decay BRs of other decay processes in several scenarios. 



This paper is organized as follows. In Sec.~\ref{sec:Model}, we define the THDMs and we mention the current situation of experimental constraints. 
In Sec.~\ref{sec:Renormalization}, we define our renormalization scheme, and give renormalized vertices $Hhh$, $Hf\bar{f}$ ($f=t,~ b,~ \tau,~ c$) and $HVV$ ($V=W,~ Z$).
In Sec.~\ref{sec:Decay_rate}, we give formulae of one-loop corrected decay rates of $H\to hh$, $H\to f\bar{f}$ and $H\to VV$.
Numerical evaluations of these decay rates are shown in Sec.~\ref{sec:Results}.
Discussions and conclusions are respectively given in Sec.~\ref{sec:Discussions} and \ref{sec:Conclusions}.
In Appendix, we present explicit analytic formulae of 1PI diagrams for the $Hhh$ vertex.

\section{Model}\label{sec:Model}
The Higgs sector is composed of two isospin doublet fields $\Phi_1^{}$ and $\Phi_2^{}$ with hypercharge $Y=1/2$. 
In order to avoid flavor changing neutral currents mediated by Higgs bosons at tree level,   
we impose a discrete $Z_2$ symmetry to the model, where $\Phi_1$ and $\Phi_2$ are transformed to $+\Phi_1$ and $-\Phi_2$~\cite{Glashow:1976nt,Paschos:1976ay}, respectively.

The Higgs potential is given under the $Z_2^{}$ symmetry by 
\begin{align}
  V &= m_1^2|\Phi_1|^2 +m_2^2|\Phi_2|^2 -m_3^2(\Phi_1^\dagger \Phi_2 +\textrm{h.c.})\notag\\
  &+ \frac{\lambda_1}{2}|\Phi_1|^4 + \frac{\lambda_2}{2}|\Phi_2|^4 + \lambda_3|\Phi_1|^2|\Phi_2|^2
  +\lambda_4 |\Phi_1^\dagger \Phi_2|^2 +\frac{\lambda_5}{2}\left[(\Phi_1^\dagger \Phi_2)^2 + \textrm{h.c.}\right],
  \label{eq:potential}
\end{align}
where the $m_3^2$ term softly breaks the $Z_2$ symmetry. 
In general, $m_3^2$ and $\lambda_5$ are complex, but we assume these parameters to be real. 
The Higgs fields are parameterized as
\begin{align}
	\Phi_i^{}=
	\begin{pmatrix}
 		w_i^+\\
 		\frac{1}{\sqrt{2}}(h_i+v_i+ iz_i)
	\end{pmatrix}, 
 \,\,\, \left(i=1, 2\right), 
\end{align}
where $v_1$ and $v_2$ are the vacuum expectation values (VEVs) with $v=\sqrt{v_1^2 +v_2^2} \simeq 246$ GeV, 
and their ratio is parameterized as $\tan\beta=v_2/v_1$. 
The mass eigenstates of the Higgs bosons are defined as follows,  
\begin{align}
  \begin{pmatrix}
    w_1^\pm\\
    w_2^\pm
  \end{pmatrix} = R(\beta)\begin{pmatrix}
    G^\pm \\
    H^\pm
  \end{pmatrix}, \,\,\,
  \begin{pmatrix}
    z_1 \\
    z_2
  \end{pmatrix} = R(\beta)\begin{pmatrix}
    G^0 \\
    A
  \end{pmatrix}, \,\,\,
  \begin{pmatrix}
    h_1 \\
    h_2
  \end{pmatrix} = R(\alpha)
  \begin{pmatrix}
    H \\
    h
  \end{pmatrix},
  \label{eq:mass_matrix}
  \end{align}
where
\begin{align}
  R(\theta) = \begin{pmatrix}
    c_\theta & -s_\theta\\
    s_\theta & c_\theta
  \end{pmatrix}, 
  \end{align}
with $s_\theta\equiv \sin\theta$ and $c_\theta\equiv \cos\theta$. 
In Eq.~(\ref{eq:mass_matrix}), $H^\pm$, $A$, $H$ and $h$ are the physical mass eigenstates, while
$G^\pm$ and $G^0$ are the Nambu-Goldstone bosons. 
After imposing the tadpole conditions, the masses of the charged and CP-odd Higgs bosons are expressed as
\begin{align}
  m_{H^\pm}^2 = M^2 -\frac{v^2}{2}(\lambda_4 +\lambda_5), \,\,\,m_A^2 =M^2 -v^2\lambda_5,
\end{align}
where $M^2$ is defined as $M^2=m_3^2/(s_\beta c_\beta)$. 
The masses of the CP-even Higgs bosons and the mixing angle $\beta-\alpha$ are given by  
\begin{align}
  &m_H^2 = M_{11}^2 c_{\beta-\alpha}^2 +M_{22}^2 s_{\beta-\alpha}^2 -M_{12}^2 s_{2(\beta-\alpha)}^{}, \\
  &m_h^2 = M_{11}^2 s_{\beta-\alpha}^2 +M_{22}^2 c_{\beta-\alpha}^2 +M_{12}^2 s_{2(\beta-\alpha)}^{}, \\  
  &\tan2(\beta-\alpha)= -\frac{2M_{12}^2}{M_{11}^2 -M_{22}^2},
\end{align}
where
\begin{align}
  M_{11}^2 &= v^2\left[\lambda_1 c_\beta^4 +\lambda_2 s_\beta^4 +2(\lambda_3 + \lambda_4 +\lambda_5)s_\beta^2 c_\beta^2\right],\\
  M_{22}^2 &= M^2 +\frac{v^2}{4}s_{2\beta}^2\left[\lambda_1 +\lambda_2 -2(\lambda_3 + \lambda_4 +\lambda_5)\right],\\
  M_{12}^2 &= \frac{v^2}{2}s_{2\beta}^{}\left[-\lambda_1 c_\beta^2 +\lambda_2 s_\beta^2 +(\lambda_3 + \lambda_4 +\lambda_5)c_{2\beta}^{}\right],
\end{align}
with $M_{ij}^2$ being elements of the mass matrix in the basis of $R(\beta)(h_1, h_2)^T$. 
In the following, we identify $h$ as the discovered Higgs boson with a mass of 125 GeV.  
The 8 parameters in Eq.~(\ref{eq:potential}) can be expressed by the following 8 parameters,
 \begin{align}
 m_h^2,~ m_H^2,~ m_A^2,~ m_{H^\pm}^2,~ M^2,~ s_{\beta-\alpha}^{},~ \tan\beta,~ v,  
 \end{align}
 where we define $\tan\beta > 0$ and $0\leq s_{\beta-\alpha}^{} \leq 1$, while $c_{\beta-\alpha}$ can be either positive and negative.  
These parameters can be constrained by considering bounds from perturbative unitarity~\cite{Kanemura:2015ska,Kanemura:1993hm,Akeroyd:2000wc,Ginzburg:2005dt} and vacuum stability~\cite{Deshpande:1977rw,Klimenko:1984qx,Sher:1988mj,Nie:1998yn}. 
Throughout the paper, we take $M^2 \geq 0$ by which the true vacuum condition is satisfied~\cite{Barroso:2013awa}.

For the later discussion, we here give expressions of the triple scalar couplings as
\begin{align}
  \lambda_{hhh}^{} &= -\frac{m_h^2}{2v}s_{\beta-\alpha}^{}
  + \frac{M^2 - m_h^2}{v}c_{\beta-\alpha}^2 \left[ s_{\beta-\alpha}^{} +\frac{c_{\beta-\alpha}}{2} (\cot\beta-\tan\beta) \right], \label{eq:lambda_hhh}\\
 \lambda_{Hhh}^{} &= - \frac{c_{\beta-\alpha}^{}}{2vs_{2\beta}^{}} \left[
 (2m_h^2 +m_H^2 -3M^2)s_{2\alpha}^{} +M^2 s_{2\beta}^{} \right],
 \label{eq:lambda_Hhh}\\
 \lambda_{HHh}^{}&= \frac{s_{\beta-\alpha}^{}}{2v}[
   (2M^2 -2m_H^2 -m_h^2)s_{\beta-\alpha}^2 +2(3M^2 -2m_H^2 -m_h^2)\cot2\beta s_{\beta-\alpha}^{}c_{\beta-\alpha}^{} \notag\\
   &-(4M^2 -2m_H^2 -m_h^2)c_{\beta-\alpha}^2]
 \label{eq:lambda_HHh}, 
\end{align}
where they are defined as $\mathcal{L} = \lambda_{\phi\phi'\phi''}\phi\phi'\phi'' + \cdots$.

The kinetic term of the Higgs fields is
\begin{align}
  \mathcal{L}_\mathrm{kin}= |D_\mu\Phi_1|^2 +|D_\mu\Phi_2|^2, 
\end{align}
where $D_\mu$ is the covariant derivative given by $D_\mu = \partial_\mu -i\frac{g}{2}\tau^aW_\mu^a -i\frac{g'}{2}B_\mu$. 
The gauge-gauge-scalar interaction terms in the mass eigenstates are extracted as
\begin{align}
  \mathcal{L}_\mathrm{kin}&\supset{g} 
  m_W^{} g^{\mu\nu}\left( \kappa_W^{h} W^{+}_\mu W^{-}_\nu h + \kappa_W^{H} W^{+}_\mu W^{-}_\nu H \right) 
  +\frac{g_Z^{} m_Z^{}}{2}g^{\mu\nu}\left(\kappa_Z^{h} Z^{}_\mu Z^{}_\nu h + \kappa_Z^{H} Z^{}_\mu Z^{}_\nu H \right), 
\end{align}
where $g_Z$ is defined as $g_Z\equiv g/\cos\theta_W$, and $\kappa_V^{\phi}$ ($\phi =h, H$) is the scaling factor obtained as
\begin{align}
  \kappa_V^{h} = s_{\beta-\alpha}^{}, \,\,\,   \kappa_V^{H} = c_{\beta-\alpha}^{}.
  \label{eq:kappa_V}
  \end{align}


\begin{table}[t]
\begin{center}
\begin{tabular}{c|cccccccccc}
\hline 
& \quad $\Phi_1$ & \quad $\Phi_2$ & \quad $Q_L$
& \quad $L_L$ & \quad $u_R$ & \quad $d_R$ & \quad $e_R$ 
& \quad $\zeta_u$ & \quad $\zeta_d$ & \quad $\zeta_e$ \\ \hline \hline
Type-I \quad & \quad $+$ & \quad$-$ & \quad$+$ & \quad$+$ & \quad$-$ & \quad$-$ & \quad$-$
& \quad$\cot\beta$ & \quad$\cot\beta$ & \quad$\cot\beta$   \\
Type-II \quad & \quad $+$ & \quad$-$ & \quad$+$ & \quad$+$ & \quad$-$ & \quad$+$ & \quad$+$
& \quad$\cot\beta$ & \quad$-\tan\beta$ & \quad$-\tan\beta$   \\
Type-X \quad& \quad $+$ & \quad$-$ & \quad$+$ & \quad$+$ & \quad$-$ & \quad$-$ & \quad$+$
& \quad$\cot\beta$ & \quad$\cot\beta$ & \quad$-\tan\beta$   \\
Type-Y \quad & \quad $+$ & \quad$-$ & \quad$+$ & \quad$+$ & \quad$-$ & \quad$+$ & \quad$-$
& \quad$\cot\beta$ & \quad$-\tan\beta$ & \quad$\cot\beta$   \\
\hline
\end{tabular}
\end{center}
\caption{Charge assignment of the $Z_2$ symmetry and mixing factors $\zeta_f^{}$ ($f=u, d, e$) in each type of the Yukawa interaction.} 
\label{tab:Yukawa_Couplings}
\end{table}

The most general Yukawa interaction is given under the $Z_2^{}$ symmetry as 
\begin{align}
  \mathcal{L}_Y^{}= -Y_u\bar{Q}_L^{}(i \tau_2 \Phi_i^*) u_R^{} -Y_d\bar{Q}_L^{}\Phi_j d_R^{} -Y_e\bar{L}_L^{}\Phi_k e_R^{} + \rm{h.c.}, 
  \end{align}
where $\Phi_i$, $\Phi_j$ and $\Phi_k$ are either $\Phi_1$ or $\Phi_2$. 
These labels of the Higgs doublet fields are determined by fixing the $Z_2$ charge of each field as summarized in Tab.~\ref{tab:Yukawa_Couplings}~\cite{Barger:1989fj,Aoki:2009ha}. 
The Yukawa interaction terms are given in terms of the mass eigenstates of the Higgs bosons as 
\begin{align}
  \mathcal{L}_Y &\supset - \sum_{f=u,d,e} \frac{m_f}{v}\left( \kappa_f^h\bar{f}fh +\kappa_f^H\bar{f}fH + \kappa_f^A\bar{f}\gamma_5^{}fA 
    \right)\notag\\
  &-\frac{\sqrt{2}}{v}\left[\bar{u}V_{ud}^{}(m_d \zeta_d P_R -m_u \zeta_u P_L)dH^+
    +m_e \zeta_e \bar{\nu} P_R eH^+ +\rm{h.c.}\right],
  \label{eq:Yukawa}
\end{align}
where $I_f^{}= 1/2$ $(-1/2)$ for $f=u$ $(d, e)$, and $V_{ud}$ is the Cabibbo-Kobayashi-Maskawa matrix element.
Scaling factors in Eq.~(\ref{eq:Yukawa}) are expressed as, 
\begin{align}
  \kappa_f^h = s_{\beta-\alpha}^{} +\zeta_f c_{\beta-\alpha}^{},
  \quad \kappa_f^H = c_{\beta-\alpha}^{} -\zeta_f s_{\beta-\alpha}^{},
  \quad \kappa_f^A = -2iI_f \zeta_f, 
  \label{eq:kappa_f}
\end{align}
where $\zeta_f$ are given in Tab.~\ref{tab:Yukawa_Couplings}.

In the limit of $s_{\beta-\alpha}^{}\to 1$, all couplings of $h$ become the same as those of the SM at tree level. 
We call this limit as the alignment limit. 
As mentioned in Sec.~\ref{sec:Intro}, we are interested in the nearly alignment case, so that 
we introduce a parameter $x$ defined by 
\begin{align}
 x \equiv - (\beta-\alpha) + \frac{\pi}{2},  
\end{align}
where the limit $x\to 0$ corresponds to the alignment limit. 
In the nearly alignment case; i.e., $x\simeq 0$, the scaling factors and the triple scalar couplings can be expressed as
\begin{align}
  \kappa_V^{h} &= 1-\frac{x^2}{2} + \mathcal{O}(x^3), \quad
  \kappa_V^H = x + \mathcal{O}(x^3), \\
  \kappa_f^h &=1 +\zeta_fx + \frac{x^2}{2} + \mathcal{O}(x^3), \quad
  \kappa_f^H = -\zeta_f + x + \frac{x^2}{2}\zeta_f + \mathcal{O}(x^3), \\
  \lambda_{ijk}^{} &= \lambda_{ijk, 0} + \lambda_{ijk, 1} x + \lambda_{ijk, 2}x^2 +\mathcal{O}(x^3), \label{eq:ppp_series}
\end{align}
with
\begin{align}
  &  \lambda_{hhh, 0} =-\frac{m_h^2}{2v}, \quad \lambda_{hhh, 1} = 0,\quad \lambda_{hhh, 2}=\frac{4M^2-3m_h^2}{4v}, \label{eq:hhh_series}\\
  & \lambda_{Hhh, 0} =0,\quad \lambda_{Hhh, 1} = \frac{2m_h^2+m_H^2-4M^2}{2v}, \quad \lambda_{Hhh, 2} =\frac{(2m_h^2 + m_H^2 -3M^2)\cot2\beta}{v}, \label{eq:Hhh_series} \\
  &  \lambda_{HHh, 0} =-\frac{m_h^2+2m_H^2 -2M^2}{2v}, \quad \lambda_{HHh, 1} = -\frac{(m_h^2 +2m_H^2-3M^2)\cot 2\beta}{v}, \notag\\
  & \lambda_{HHh, 2} =\frac{5m_h^2+10m_H^2 -14M^2}{4v}.     \label{eq:HHh_series}
\end{align}


We mention constraints from current experimental data on the THDM. \\
\begin{itemize}
\item Electroweak precision data\\
New physics effects on the EW oblique parameters are expressed by the $S$, $T$ and $U$ parameters~\cite{Peskin:1990zt,Peskin:1991sw}.  
It has been known that the $T$ parameter represents the violation of the custodial $SU(2)_V$ symmetry~\cite{Haber:1992py,Pomarol:1993mu,Herquet:2008eaa,Kanemura:2011sj}.
One of the ways to restore the $SU(2)_V$ symmetry in the potential is to take $m_{H^\pm} =m_A$, by which quadratic-power like dependences of Higgs boson masses on the $T$ parameter disappear.~\footnote{We can impose the so-called twisted-custodial symmetry by taking $M=m_H=m_{H^\pm}$ and $s_{\beta-\alpha}=1$ ~\cite{Gerard:2007kn,Aiko:2020atr} instead of $m_A=m_{H^\pm}$. In order to realize $\Delta T=0$ at one-loop level, however, it is enough to take $m_{H^\pm}=m_H$ and $s_{\beta-\alpha}=1$.  } 

By using the global fit of EW parameters~\cite{Baak:2012kk}, new physics effects on the $S$ and $T$ parameters under $U=0$ are constrained by
\begin{align}
  S = 0.05\pm0.09,\quad T = 0.08\pm0.07,
\end{align}
with the correlation factor of $+0.91$~\cite{Baak:2012kk}. 

\item Signal strength of the Higgs boson $h$\\
Measurements of the signal strengths of $h$ constrain mixing parameters $\alpha$ and $\beta$ 
as seen in Eqs.~(\ref{eq:kappa_V}) and (\ref{eq:kappa_f}). 
According to Refs.~\cite{CMS:2020gsy,ATLAS:2021vrm}, the parameter regions with $-0.25 \lesssim c_{\beta-\alpha}^{} \lesssim 0.01$ for $\tan\beta \simeq 2$ are allowed in the Type-I THDM. 
In the Type-II, X and Y THDMs, the constraint is given by; e.g., $|c_{\beta-\alpha}| \lesssim 0.1$ for $\tan\beta \simeq 2$, and stronger bounds are taken for larger $\tan\beta$. 
\item Direct searches for the additional Higgs bosons at LHC \\
  So far, there has been no report for the discovery of the additional Higgs bosons at LHC, so that lower limits on their masses have been taken. 
In the THDMs, constraints from the direct searches have been studied by using LHC Run-II data in; e.g., Refs.~\cite{Aiko:2020ksl,Kling:2020hmi,Cheung:2022ndq}.   
In the nearly alignment region, the search for the $pp\to A\to Zh$ process typically excludes the largest region of the parameter space in the four types of THDMs. 
For instance,  $m_A \lesssim 2m_t$ (900 GeV) are excluded in the case with $s_{\beta-\alpha}^{} = 0.99$ and $\tan\beta \simeq 5$ (2)\cite{Aiko:2020ksl}. 
For larger values of $1-s_{\beta-\alpha}^{}$, the constraint by this mode becomes stronger. 
In Type-II and Type-Y, the search for the $pp\to b\bar{b}A\to b\bar{b}Zh$ process also provides severe constraints on the parameter space particularly for the case with larger $\tan\beta$ values. 
For instance, in the case with $s_{\beta-\alpha} = 0.99$ and $\tan\beta = 10$, $m_A\lesssim 500$ GeV have been excluded~\cite{Aiko:2020ksl}. 
In addition to the $A\to Zh$ mode, the $H\to hh$ mode can also exclude wide regions of the parameter space.
This, however, strongly depends on the value of $M^2$, and can be changed by the radiative corrections which will be discussed later.    
\item Flavor experiments\\
The mass of the charged Higgs bosons is restricted by the $B\to X_s\gamma$ decay. 
In Type-II and Type-Y, $m_{H^{\pm}}^{}\lesssim 800$ GeV is excluded for $\tan\beta > 2$~\cite{Misiak:2017bgg, Misiak:2020vlo}.  
On the other hand, in Type-I and Type-X, $m_{H^\pm}^{} \lesssim 400$ GeV is excluded for $\tan\beta=1$~\cite{Misiak:2017bgg, Misiak:2020vlo}.  
Moreover, in the Type-II case, data of the $B_s\to \mu\mu$ decay exclude regions with large $\tan\beta$~\cite{Cheng:2015yfu,Haller:2018nnx,Enomoto:2015wbn}; e.g., $\tan\beta \gtrsim 25$ for $m_{H^\pm}=1$ TeV.   
\end{itemize}

\section{Renormalized vertices}\label{sec:Renormalization}
In this section, we discuss the renormalized vertices for $H$ based on the improved on-shell scheme~\cite{Kanemura:2017wtm},  
where gauge dependences in counter terms of mixing angles are removed by adding pinch terms. 
This treatment is not applied to wave function renormalizations but counter terms from shifts of parameters in the Lagrangian.

In the following discussion, we decompose each renormalized form factor $\hat{\Gamma}_{HXX}^{}$ ($X=h, W, Z, t, b, \tau, c$) as 
\begin{align}
  \hat{\Gamma}_{HXX}^{} &= \Gamma_{HXX}^\textrm{Tree} + \Gamma_{HXX}^\textrm{1-loop},
\end{align}
where
\begin{align}
 \Gamma_{HXX}^\textrm{1-loop} = \delta \Gamma_{HXX}^{} + \Gamma_{HXX}^\textrm{1PI} 
 + \Gamma_{HXX}^\textrm{Tad},
 \label{eq:Gamma_1-loop}
\end{align}
with $\Gamma_{HXX}^\textrm{Tree}$, $\delta\Gamma_{HXX}^{}$, $\Gamma_{HXX}^\textrm{1PI}$ and $\Gamma_{HXX}^{\textrm{Tad}}$ 
being contributions from tree level diagrams, counter terms, 1PI diagrams and tadpole diagrams inserted to tree level vertices, respectively.

\subsection{Renormalization conditions}
We discuss renormalization conditions in order to determine the counter terms for the renormalized $HXX$ vertices. 
The renormalized two-point functions for neutral scalar bosons are expressed as 
 \begin{align}
   &\hat{\Pi}_{ii}[p^2] =(p^2-m_i^2)(1+\delta Z_i) -\delta m_i^2 + \tilde{\Pi}_{ii}^\mathrm{1PI}[p^2] , \quad (i=h, H, A) \label{eq:reno-two-func1}  \\ 
   &\hat{\Pi}_{hH}[p^2] =p^2(\delta C_{hH}+\delta C_{Hh}) + m_h^2(\delta\alpha_f -\delta C_{hH}^{}) - m_{H}^2(\delta\alpha_f + \delta C_{Hh}^{}) + \tilde{\Pi}_{hH}^\mathrm{1PI}[p^2] , \label{eq:reno-two-func2}  \\ 
   &\hat{\Pi}_{AG^0}[p^2] =p^2(\delta C_{AG^0}+\delta C_{G^0A}) +m_{A}^2(\delta\beta_f - \delta C_{AG^0}^{})+\tilde{\Pi}_{AG^0}^\mathrm{1PI}[p^2] , \label{eq:reno-two-func3}  
 \end{align}
 where $\delta Z_i$, $\delta C_{ij}$, $\delta\alpha_f$ and $\delta\beta_f$ come from the field shifts defined as
 \begin{align}
   &\begin{pmatrix}
     H\\
     h\\
   \end{pmatrix} \to \begin{pmatrix}
     1+\frac{1}{2}\delta Z_H & \delta C_{Hh} +\delta\alpha_f \\
     \delta C_{hH} -\delta\alpha_f & 1+\frac{1}{2}\delta Z_h \\
   \end{pmatrix} \begin{pmatrix}
     H\\
     h\\
     \end{pmatrix}, \\
  &\begin{pmatrix}
     G^0\\
     A\\
   \end{pmatrix} \to \begin{pmatrix}
     1+\frac{1}{2}\delta Z_{G^0} & \delta C_{G^0A} +\delta\beta_f \\
     \delta C_{AG^0} -\delta\beta_f & 1+\frac{1}{2}\delta Z_A \\
   \end{pmatrix} \begin{pmatrix}
     G^0\\
     A\\
     \end{pmatrix}. 
 \end{align} 
 We here distinguish $\delta\alpha_f$ ($\delta\beta_f$) from $\delta\alpha$ ($\delta\beta$) which appears from the parameter shift; i.e., $\alpha\to\alpha +\delta\alpha$ ($\beta\to\beta +\delta\beta$). 
 In Eqs.~(\ref{eq:reno-two-func1})-(\ref{eq:reno-two-func3}), $\tilde{\Pi}_{ij}^\textrm{1PI}$ is defined as 
 \begin{align}
   \tilde{\Pi}_{ij}^\textrm{1PI}[p^2]= \Pi_{ij}^\textrm{1PI}[p^2]+ \Pi_{ij}^\textrm{Tad}. 
 \end{align}

 By imposing the following on-shell conditions for the above renormalized two-point functions; 
 \begin{align}
   \hat{\Pi}_{ii}[m_i^2]=0, \quad \frac{d}{dp^2}\hat{\Pi}_{ii}^{}[p^2]\big|_{p^2=m_i^2} = 1,  \,\,\,(i=h, H, A), \quad
   \end{align}
 the counter terms $\delta m_i^2$ and $\delta Z_i$ 
 are determined as 
 \begin{align}
   &\delta m_i^2 = \tilde{\Pi}_{ii}^\textrm{1PI}[m_i^2], \quad
   \delta Z_i = -\frac{d}{dp^2}\Pi_{ii}^\textrm{1PI}[p^2]\big|_{p^2=m_i^2}. 
   \end{align}

 By imposing the on-shell conditions for the mixing two-point functions as 
 \begin{align}
   &\hat{\Pi}_{hH}[m_h^2]=\hat{\Pi}_{hH}[m_H^2]=0, 
   \quad \hat{\Pi}_{AG^0}[m_A^2]=\hat{\Pi}_{AG^0}[0]=0, 
 \end{align}
we obtain 
\begin{align}
  &\delta C_h^{} =\frac{1}{2(m_H^2-m_h^2)}\left(\Pi_{Hh}^\textrm{1PI}[m_h^2] - \Pi_{Hh}^\textrm{1PI}[m_H^2]\right),\\
  &\delta C_A^{} =-\frac{1}{2m_A^2}\left(\Pi_{Hh}^\textrm{1PI}[m_A^2] - \Pi_{Hh}^\textrm{1PI}[0]\right),\\
  &\delta \alpha_f =\frac{1}{2(m_H^2-m_h^2)}\left(\tilde{\Pi}_{Hh}^\textrm{1PI}[m_h^2]  + \tilde{\Pi}_{Hh}^\textrm{1PI}[m_H^2] \right), \\
 &\delta \beta_f = -\frac{1}{2m_A^2}\left(\tilde{\Pi}_{AG}^\textrm{1PI}[m_A^2]  + \tilde{\Pi}_{AG}^\textrm{1PI}[0] \right),    
\end{align}
where we take $\delta C_{hH} = \delta C_{Hh} \equiv \delta C_h $ and $\delta C_{AG^0} = \delta C_{G^0A} \equiv \delta C_A$. 

As mentioned above, the counter terms $\delta\alpha$ and $\delta\beta$ should include the pinch term; 
\begin{align}
  &\delta \alpha =\frac{1}{2(m_H^2-m_h^2)}\left(\tilde{\Pi}_{Hh}^\textrm{1PI}[m_h^2] +\Pi_{Hh}^\textrm{PT}[m_h^2] + \tilde{\Pi}_{Hh}^\textrm{1PI}[m_H^2] + \Pi_{Hh}^\textrm{PT}(m_H^2) \right),\\
  &\delta \beta = -\frac{1}{2m_A^2}\left(\tilde{\Pi}_{AG}^\textrm{1PI}[m_A^2]  +\Pi_{AG}^\textrm{PT}[m_A^2]+ \tilde{\Pi}_{AG}^\textrm{1PI}[0] +\Pi_{AG}^\textrm{PT}[0] \right), 
  \end{align}
where explicit formulae of $\Pi_{Hh}^\textrm{PT}(p^2)$ and $\Pi_{AG}^\textrm{PT}(p^2)$ are given in Ref.~\cite{Kanemura:2017wtm}.

The counter term $\delta M^2$ cannot be determined by using the on-shell conditions discussed above.
Thus, we apply the $\overline{\textrm{MS}}$ scheme to the renormalization of the $hhh$ vertex, 
and then $\delta M^2$ is determined as 
\begin{align}
&\delta M^2 =\frac{M^2}{16\pi^2v^2}\left[2\sum_fN_c^f m_f^2 \kappa_f^2 + 4M^2 -2m_{H^\pm}^2 -m_A^2 + \frac{s_{2\alpha}}{s_{2\beta}}(m_H^2 - m_h^2)-3(2m_W^2 + m_Z^2)\right]\Delta_\textrm{Div} \notag\\
 & + \frac{2c_{2\beta}^{}}{s_{2\beta}^{}}\frac{M^2}{v} 
  \left(\frac{c_{\beta-\alpha}^{}}{m_h^2} \textrm{Div(}T_{h}^\textrm{1PI}) 
  - \frac{s_{\beta-\alpha}^{}}{m_H^2} \textrm{Div}(T_{H}^\textrm{1PI})\right),  \label{eq:deltaM2}
\end{align}
where $\Delta_\textrm{Div} = 1/\epsilon -\gamma_E + \log 4\pi$ with $\gamma_E$ being the Euler's constant.
In the second line, Div$(T_{h(H)}^\textrm{1PI})$ indicates the divergent part of tadpole type diagrams whose explicit formulae are
given in Ref.~\cite{Kanemura:2015mxa}. 
In this prescription, renormalized quantities including $\delta M^2$ such as the decay rate of $H\to hh$ have a dependence on the renormalization scale $\mu$.

Finally, the counter terms of the EW parameters $\delta v$, $\delta m_V^2$, $\delta Z_V$, $\delta m_f$ and $\delta Z_V^f$ are given in Ref.~\cite{Kanemura:2015mxa}. 

\subsection{Renormalized $Hhh$ vertex}\label{sec:Hhh}
The contributions from the tree level diagram and from counter terms are given by
\begin{align}
  &\Gamma_{Hhh}^\textrm{Tree} = 2\lambda_{Hhh}^{}, \\
  &\delta \Gamma_{Hhh}^{} =
  2\delta\lambda_{Hhh}^{} +\lambda_{Hhh}^{}( \delta Z_H^{} +2\delta Z_h^{}) 
 + 6 \lambda_{hhh}^{} (\delta C_h^{} - \delta\alpha_f) + 4\lambda_{HHh}^{} (\delta C_h^{} + \delta\alpha_f), 
\end{align}
where $\lambda_{hhh}^{}$, $\lambda_{Hhh}$ and $\lambda_{HHh}$ are given in Eqs.~(\ref{eq:lambda_hhh})-(\ref{eq:lambda_HHh}),
and $\delta\lambda_{Hhh}$ is given by
\begin{align}
 \delta\lambda_{Hhh}^{} & =
 -\lambda_{Hhh}^{}\frac{\delta v}{v} -\frac{s_{2\alpha}^{}c_{\beta -\alpha}^{}}{2vs_{2\beta}^{}}(2 \delta m_h^2 + \delta m_H^2) 
 +G_{\alpha}^{}\delta\alpha + G_\beta^{}\delta\beta 
 + \frac{3c_{\beta-\alpha}^{} }{2v}\left(\frac{s_{2\alpha}^{}}{s_{2\beta}^{}} - \frac{1}{3}\right)\delta M^2, 
\end{align}
with $G_\alpha^{}$ and $G_\beta^{}$ being 
\begin{align}
 G_\alpha^{} &= \tan(\beta -\alpha)\lambda_{Hhh}^{} 
 - \frac{c_{2\alpha}^{}c_{\beta -\alpha}^{}}{vs_{2\beta}^{}}(2m_h^2 +m_H^2 -3M^2), \label{eq:Galpha}\\
 G_\beta^{} &= -\frac{c_{\alpha-3\beta}^{}+3c_{\alpha+\beta}^{}}{2s_{2\beta}^{}c_{\beta-\alpha}^{}}\lambda_{Hhh}^{}
 - \frac{c_{\beta-\alpha}^{}}{v}\cot2\beta M^2.
 \label{eq:Gbeta}
\end{align}
Explicit formulae of contributions of 1PI diagrams $\Gamma_{Hhh}^\textrm{1PI}$ are given in Appendix. 
As mentioned in the previous subsection, the $Hhh$ coupling depends on the renormalization scale $\mu$.

\subsection{Renormalized $Hf\bar{f}$ vertex}\label{sec:hff}
The renormalized $H f\bar{f}$ ($f= t,~ b,~ c,~ \tau$) vertex is expressed in terms of the following form factors, 
\begin{align}
 \hat{\Gamma}_{Hff}^{}[p_1^2, p_2^2, q^2] &= 
 \hat{\Gamma}_{Hff}^{S} + \gamma_5 \hat{\Gamma}_{Hff}^{P} + \sum_{i=1,2}\left(\Slash{p}_i^{}\hat{\Gamma}_{Hff}^{V_i^{}}
 + \Slash{p}_i^{}\gamma_5^{}\hat{\Gamma}_{Hff}^{A_i^{}} \right)
 + \Slash{p}_1^{}\Slash{p}_2^{}\hat{\Gamma}_{Hff}^{T}
 + \Slash{p}_1^{}\Slash{p}_2^{}\gamma_5^{}\hat{\Gamma}_{Hff}^{PT}, 
\end{align} 
where $p_1^\mu$ and $p_2^\mu$ are the incoming momenta of external particles $f$ and $\bar{f}$, respectively, and 
$q^\mu (=p_1^\mu + p_2^\mu)$ is the outgoing momentum of $H$. 
For the case with on-shell fermions; i.e., $p_1^2 =p_2^2=m_f^2$, the following relations hold: 
\begin{align}
 \hat{\Gamma}_{Hff}^{P} = \hat{\Gamma}_{Hff}^{PT} =0, \hspace{0.5cm}
 \hat{\Gamma}_{Hff}^{V_1^{}} = -\hat{\Gamma}_{Hff}^{V_2^{}} , \hspace{0.5cm}
 \hat{\Gamma}_{Hff}^{A_1^{}} = -\hat{\Gamma}_{Hff}^{A_2^{}}. \label{eq:hff_form_relation}
\end{align}
The contributions from the tree level diagram and from counter terms are given by 
\begin{align} 
 &\Gamma_{Hff}^{\textrm{Tree}, S} = - \frac{m_f^{}}{v}\kappa^H_{f}, \quad \Gamma_{Hff}^{\textrm{Tree}, i} = 0, \\
 &\delta\Gamma_{Hff}^{S} = -\frac{m_f^{}}{v}\kappa^{H}_f\left[
  \frac{\delta m_f^{}}{m_f^{}} - \frac{\delta v}{v} + \frac{1}{2}\delta Z_H + \delta Z_V^f 
  + \frac{\delta\kappa^H_f}{\kappa^H_f} + \frac{\kappa^h_f}{\kappa^H_f}(\delta C_h - \delta\alpha_f^{})\right], 
 \quad \delta\Gamma_{Hff}^i =0, 
\end{align}
where the index $i$ runs over $i=\{P, \,V_1,\, V_2,\, A_1,\,A_2,\, T,\,PT\}$, and 
\begin{align}
&\delta\kappa^H_f = \kappa_f^h\delta\alpha -\kappa_f^H \zeta_f\delta\beta.  
\end{align}
The mixing factor $\zeta_f^{}$ and $\kappa_f^\phi$ ($\phi =h$, $H$) are given in Sec.~\ref{sec:Model}.

\subsection{Renormalized $HVV$ vertex}
The renormalized $HV^\mu V^\nu$ ($V =W, Z$) vertex is composed of three types of form factors expressed as 
\begin{align} 
 \hat{\Gamma}_{HVV}^{\mu\nu}[p_1^2, p_2^2, q^2] = 
 g^{\mu\nu}\hat{\Gamma}_{HVV}^{1} + \frac{p_1^\mu p_2^\nu}{m_V^2}\hat{\Gamma}_{HVV}^{2} 
 + i\epsilon^{\mu\nu\rho\sigma}\frac{p_{1, \rho}p_{2, \sigma}}{m_V^2}\hat{\Gamma}_{HVV}^{3}, 
\end{align}
 where $p_1^\mu$ and $p_2^\nu$ are incoming momenta of the weak bosons, and $q^\mu$ is the outgoing momentum of $H$.

The contributions from the tree level diagram and from counter terms are given by 
\begin{align} 
 &\Gamma_{HVV}^{\textrm{Tree}, 1} = \frac{2m_V^2}{v} c_{\beta-\alpha}^{}, 
 \quad \Gamma_{HVV}^{\textrm{Tree}, 2} = \Gamma_{HVV}^{\textrm{Tree}, 3} = 0, \label{eq:HVV_tree} \hspace{0.5cm}\\
 &\delta\Gamma_{HVV}^{1} = \frac{2m_V^2}{v}c_{\beta-\alpha}^{} \left[
 \frac{\delta m_V^2}{m_V^2} -\frac{\delta v}{v} +\delta Z_V^{} + \frac{1}{2}\delta Z_H^{} 
 + \tan(\beta- \alpha)\left(\delta\alpha - \delta\beta + \delta C_h^{} -\delta\alpha_f^{}\right) 
 \right], \\
 &\delta\Gamma_{HVV}^{2} = \delta\Gamma_{HVV}^{3} = 0. \label{eq:HVV_counterterm}
\end{align}

\section{Radiative corrections to decay rates}\label{sec:Decay_rate}
In this section, we give formulae of the decay rates of $H$ with NLO corrections in EW and scalar interactions.
In particular, we focus on the processes $H\to hh$, $H\to f\bar{f}$ and $H\to VV$ and the case where $H$ is lighter than $A$ and $H^\pm$, 
so that the decays of $H\to AZ$, $H\to H^\pm W$, $H\to H^+ H^-$ and $H\to AA$ are kinematically forbidden.
For $H\to q\bar{q}$ and loop induced processes, we implement QCD corrections to their decay rates. 
The squared amplitude is given at NLO in EW and scalar interactions by 
\begin{align}
  \left[\left|\mathcal{M}[H\to XX]\right|^2\right]_\textrm{NLO} = \left|\mathcal{M}_\textrm{Tree}\right|^2
  + \left( \mathcal{M}_\textrm{1-loop}^\dagger \mathcal{M}_\textrm{Tree}+ \textrm{h.c.} \right).
  \label{eq:amplitude}
\end{align}
Since we drop the one-loop squared term $|\mathcal{M}_{\textrm{1-loop}}^{}|^2$ which corresponds to next-to-NLO (NNLO) corrections, the decay rates can be negative values depending on the parameter choice. 
Although we can avoid such a strange behavior by adding $|\mathcal{M}_{\textrm{1-loop}}^{}|^2$,
we also need to include terms with two-loop diagrams multiplied by $\mathcal{M}_\textrm{Tree}$ 
for the consistent perturbative calculation at NNLO. 
We note that in the alignment limit, adding $|\mathcal{M}_{\textrm{1-loop}}^{}|^2$ to Eq.~(\ref{eq:amplitude}) would be justified because the tree level contribution vanishes. 
In the following calculation, we simply use Eq.~(\ref{eq:amplitude}), where we exclude parameter regions giving rise to negative values of decay rates.

\subsection{Decay rate of $H \to hh$}\label{sec:H to hh}
We consider the $H \to hh$ decay mode which is kinematically allowed for $m_H \geq 2m_h$. 
The partial decay width at NLO is expressed as 
\begin{align}
 \Gamma[H\to hh] = \Gamma_\textrm{LO}[H\to hh] 
  \left( 1 + \frac{1}{\lambda_{Hhh}^{}}\textrm{Re}[\Gamma_{Hhh}^\textrm{1-loop}]  -\Delta r
\right), \label{eq:decay_Htohh}
\end{align}  
where $\lambda_{Hhh}$ and $\Gamma_{Hhh}^\textrm{1-loop}$ are given in Eq.~(\ref{eq:lambda_Hhh}) and (\ref{eq:Gamma_1-loop}), respectively.  
The LO contribution $\Gamma_\textrm{LO}[H\to hh]$ is given by 
\begin{align}
 \Gamma_\textrm{LO}[H\to hh] = \frac{\lambda_{Hhh}^2}{8\pi m_H^{}} 
  \sqrt{ 1 -4\frac{m_h^2}{m_H^2}}.    \label{eq:Gamma_Htohh}
\end{align}
In Eq.~(\ref{eq:decay_Htohh}), $\Delta r$ represents EW radiative corrections to the VEV $v$, which should be added to the decay rate,
because we choose $\alpha_\textrm{em}^{}$, $m_Z^{}$ and $G_F^{}$ as the EW input parameters in the renormalization calculation. 
The analytic expression for $\Delta r$ is given by~\cite{Sirlin:1980nh} 
\begin{align}
  \Delta r= \frac{\textrm{Re}\hat{\Pi}_{WW}^{}(0)}{m_W^2} +\frac{\alpha_\textrm{em}}{4\pi s_W^2}\left(
   6+\frac{7-4s_W^2}{2s_W^2}\log c_W^2\right).  
\end{align}
The same procedure is also applied to the other decay rates given below.

\begin{figure}
 \centering
 \includegraphics[width=160mm]{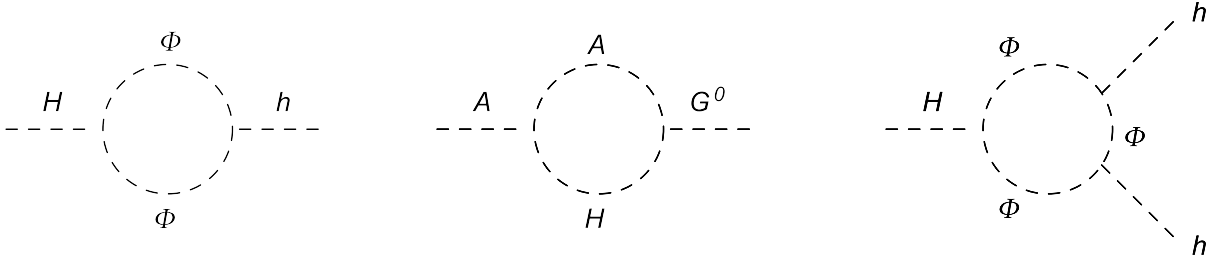}\vspace{0.5cm}
 \caption{One-loop diagrams which give major quantum effects in the alignment limit. The symbol $\Phi$ represents additional Higgs bosons; i.e. $\Phi=H, A, H^\pm$. }  
  \label{fig:diagrams_20211031} 
\end{figure}

In the nearly alignment region; i.e. $x\simeq 0$,   
the decay rate given in Eq.~(\ref{eq:decay_Htohh}) can be expanded in terms of $x$ as 
 \begin{align}
   \Gamma[H\to hh]
   &=\tilde{\Gamma}_\textrm{LO}[H\to hh]\left[
   \frac{xv}{\lambda_{Hhh, 1}^{}}\Delta_{0}^\textrm{EW}
   + x^2\left( 1  + \frac{v}{\lambda_{Hhh, 1}^{}}\Delta_{1}^\textrm{EW} + \frac{v\lambda_{Hhh, 2}}{(\lambda_{Hhh, 1})^2}\Delta_{0}^\textrm{EW} 
   -\Delta r\right) \right] +\mathcal{O}(x^3), 
   \label{eq:appro0}
\end{align}
with  
\begin{align}
  & \tilde{\Gamma}_\textrm{LO}[H\to hh] 
   \equiv \frac{1}{x^2}\Gamma_\textrm{LO}[H\to hh], \label{eq:tilde_Gam_tree}\\[8pt]
%
%
  &\frac{1}{v} \textrm{Re}[\Gamma_{Hhh}^\textrm{1-loop}]
  =\Delta_0^\textrm{EW} +x \Delta_1^\textrm{EW} + \mathcal{O}(x^2).   \hspace{0.5cm}  
\end{align}
In Eq.~(\ref{eq:appro0}), $\lambda_{Hhh ,1}^{}$ and $\lambda_{Hhh, 2}$ are given in Eq.~(\ref{eq:Hhh_series}). 
The bosonic loop and fermionic loop contributions to $\Delta_0^\textrm{EW}$ are respectively expressed as 
\begin{align}
 & \Delta_{0,B}^\textrm{EW} = \frac{2\cot 2\beta m_H^2}{16\pi^2 v^2}\left(1 - \frac{M^2}{m_H^2}\right)\Bigg\{\notag\\
    &\sum_{\Phi=H, A, H^\pm}C_\Phi^{}\lambda_{\Phi\Phi h, 0}^{}\left[ 
    \frac{1}{v}\left(B_0[m_H^2;\Phi,\Phi] +2B_0[m_h^2;\Phi,\Phi]\right)
    - 4\frac{\lambda_{\Phi\Phi h, 0}}{D_\Phi^{}}C_0[\Phi,\Phi,\Phi]\right]\notag\\
    & +\sum_{\Phi=H, A, H^\pm}\frac{C_\Phi^{}\lambda_{\Phi\Phi h, 0}^{}}{m_H^2 -m_h^2}\Bigg[
   \left( \frac{3m_h^2}{v}+\lambda_{Hhh, 1}\right)  B_0[m_H^2; \Phi,\Phi]
    +  (\lambda_{Hhh, 1}+ 4\lambda_{HHh, 0})  B_0[m_h^2; \Phi,\Phi]
     \Bigg]\notag\\
      &+\frac{\lambda_{Hhh, 1}}{vm_A^2}\Big[2(A[A]-A[H])  
    - (m_A^2-m_H^2)\left(B_0[m_A^2;H,A] +B_0[0;H,A]\right)\Big]
    \Bigg\},
    \label{eq:Delta_0_B}\\
    &  \Delta_{0,F}^\textrm{EW} = \sum_{f}\frac{2N_c^fm_f^2\zeta_f}{16\pi^2v^2}\Bigg\{
  \frac{4m_f^2}{v^2}\left[
    B_0[m_H^2,f,f] + 2B_0[m_h^2;f,f] -\frac{1}{2}(2m_h^2+m_H^2-4m_f^2)C_0[f,f,f]\right]\notag\\
  & +\frac{1}{(m_h^2-m_H^2)}\left[
    (m_H^2-4m_f^2)\left( \frac{3m_h^2}{v}+\lambda_{Hhh, 1}\right)B_0[m_H^2;f,f] 
    +\frac{m_h^2-4m_f^2}{v}(\lambda_{Hhh,1}^{} +4\lambda_{HHh,0}^{})B_0[m_h^2;f,f]\right] \notag\\
 &+\frac{\lambda_{Hhh,1 }^{}}{v}B_0[m_A^2;f,f] 
  \Bigg\},\label{eq:Delta_0_F}
\end{align}
with $N_c^f$ being the color factor and 
\begin{align}
  C_H = 3,\quad C_A=C_{H^\pm}^{}=1, \quad D_H=D_A=1,\quad D_{H^\pm}^{}=2, \quad \lambda_{\Phi\Phi h,0}^{} =-D_\Phi\frac{m_h^2 +2m_\Phi^2-2M^2}{2v}.
\end{align}
The functions $A[X]$, $B_0[q^2;X,Y]$ and $C_0[X,Y,Z](\equiv C_0[p_1^2,p_2^2,q^2; X,Y,Z])$ represent Passarino-Veltman functions~\cite{Passarino:1978jh}.
In Eq.~(\ref{eq:Delta_0_B}), terms in the second line are from $\Gamma_{Hhh}^\textrm{1PI}$, 
those in the third line are from $\delta C_h$ and $\delta \alpha$, and
those in the fourth line are from $\delta\beta$. 
In particular, for $\Delta_{0, B}^\textrm{EW}$ diagrams shown in Fig.~\ref{fig:diagrams_20211031} give the dominant contribution.  
For $\Delta_{0, F}^\textrm{EW}$, light fermion loops can be neglected because there is the overall factor $m_f^2$,
so that the top quark loop gives the dominant effect, which is proportional to $\zeta_t = \cot\beta$ in four types of Yukawa interaction. 
Therefore, the major quantum effects in the nearly alignment scenario do not depend on the types of Yukawa interaction.

Let us here consider the $x^2$ term in Eq.~(\ref{eq:appro0}), in which the first term in the parentheses is the contribution from the tree level diagram,
while the others are those from the one-loop diagrams which are suppressed by the loop factor $(1/16\pi^2)$. 
If we regard the effect of the loop suppression factor as the small expansion parameter $x$, 
the decay rate can be approximately rewritten as 
 \begin{align}
   \Gamma[H\to hh]
   &\simeq \tilde{\Gamma}_\textrm{LO}[H\to hh]\left(
   \frac{xv}{\lambda_{Hhh,1}^{}} \Delta_{0}^\textrm{EW}
   + x^2 \right). 
   \label{eq:appro1}
 \end{align}

The renormalization scale $\mu$ appears in the one-loop corrected decay rate of $H\to hh$ as 
\begin{align}
  \Gamma[H\to hh] = \Gamma_\textrm{LO}[H\to hh]\left( \Delta_{\mu}^{} +(\mu\textrm{-independent part})\right). 
\end{align}
where 
\begin{align}
  \Delta_{\mu}^{} &= -\frac{3c_{\beta-\alpha}}{v\lambda_{Hhh}^{}}\left(\frac{s_{2\alpha}}{s_{2\beta}} -\frac{1}{3}\right)
  \Bigg\{\frac{M^2}{16\pi^2v^2}\Bigg[
    2\sum_f N_c^f m_f^2 (\kappa_f^h)^2 +4M^2 -2m_{H^\pm}^2 -m_A^2 +\frac{s_{2\alpha}}{s_{2\beta}}(m_H^2-m_h^2) \notag\\
    &-3(2m_W^2 +m_Z^2)\Bigg]
    +2\cot2\beta\frac{M^2}{v}\left(\frac{c_{\beta-\alpha}}{m_h^2}T_{h}^{\mu\textrm{-part}} -\frac{s_{\beta-\alpha}}{m_H^2}T_{H}^{\mu\textrm{-part}} \right)\Bigg\}\log\mu^2. 
  \label{eq:Delta_mu}
\end{align}
The tadpole terms $T_\phi^{\mu\textrm{-part}}$ ($\phi =h$, $H$) give the dominant contribution to the $\mu$ dependence, which are proportional to  
$\cot 2\beta$, so that the magnitude of $\Delta_\mu$ grows as $\tan\beta$ becomes large.
We also mention that $\Delta_\mu$ is roughly proportional to $m_\Phi^4$, which comes from the contributions from the tadpole diagram of $h$. 
In the following, we fix as $\mu=m_H^{}$ in the numerical calculations.

\subsection{Decay rate of $H \to f\bar{f}$}
The decay rate of the process $H\to f\bar{f}$ with NLO corrections in EW and scalar interactions and QCD-NNLO corrections can be written as   
\begin{align}
\Gamma[H\to f\bar{f}] = \Gamma_\textrm{LO}^{}[H\to f\bar{f}] \left(1 + \Delta_{Hff}^\textrm{EW}  +\Delta_{Hff}^\textrm{QED}  - \Delta r+\Delta_{Hff}^\textrm{QCD}\right), \label{eq:H to ff}
\end{align}
where $\Gamma_\textrm{LO}[H\to f\bar{f}]$ is the decay rate at tree level expressed as
\begin{align}
\Gamma_\textrm{LO}^{}[H\to f\bar{f}] = \frac{N_c^f}{8\pi}m_H^{}\left(\Gamma_{Hff}^{\textrm{Tree}, S}\right)^2
 \left(1-\frac{4m_f^2}{m_H^2}\right)^{3/2}. 
\end{align}
In Eq.~(\ref{eq:H to ff}), $\Delta_{Hff}^\textrm{EW}$ represents NLO correction in EW and scalar interactions, which is expressed as 
\begin{align}
 \Delta_{Hff}^\textrm{EW}& = \frac{2}{\Gamma_{Hff}^{\textrm{Tree}, S}}\left[ 
 \textrm{Re}\left[\Gamma_{Hff}^{\textrm{1-loop}, S}\right] + 2m_f^{}\textrm{Re}\left[\Gamma_{Hff}^{\textrm{1-loop}, V_1}\right]
 + m_H^2\left(1- \frac{m_f^2}{m_H^2}\right)\textrm{Re}\left[\Gamma_{Hff}^{\textrm{1-loop}, T}\right]  
\right]. \label{eq:Delta_Hff_loop1} 
\end{align}
Contributions from virtual photon loop diagrams and real photon emissions are represented by $\Delta_{Hff}^\textrm{QED}$, 
where the term proportional to $m_f^2/m_H^2$ is neglected.  
For the leptonic decays; i.e., $f=\ell$, the QED correction calculated in the on-shell scheme is given by~\cite{Kniehl:1991ze,Dabelstein:1991ky,Bardin:1990zj}
\begin{align}
  \Delta_{H\ell\ell}^{\textrm{QED}} =\frac{\alpha_\textrm{em}^{}}{\pi}Q_\ell^2\left(\frac{9}{4}+\frac{3}{2}\log\frac{m_\ell^2}{m_H^2}\right). 
\end{align}
For the hadronic decays; i.e., $f=q$, the QED correction is given in the $\overline{\textrm{MS}}$ scheme as~\cite{Mihaila:2015lwa}
\begin{align}
  \Delta_{Hqq}^{\textrm{QED}} =\frac{\alpha_\textrm{em}^{}}{\pi}Q_q^2\left(\frac{17}{4}+\frac{3}{2}\log\frac{\mu^2}{m_H^2}\right),  
\end{align}
where we fix the renormalization scale $\mu$ as $\mu=m_H$ in numerical calculations in this paper. 

For decays into a quark pair, we implement NNLO-QCD corrections in the $\overline{\textrm{MS}}$ scheme according to
the formulae summarized in Ref.~\cite{Aiko:2020ksl}. 

\subsection{Decay rate of $H \to VV$}
We give the formulae of decay rates into a pair of on-shell gauge bosons; i.e., $H \to VV$, with NLO corrections in EW and scalar interactions. 
We express the decay rates as 
\begin{align}
  \Gamma[H\to ZZ]& = \Gamma_\textrm{LO}[H\to ZZ]\left( 1  +\Delta_{HZZ}^\textrm{EW} - \Delta r
  \right), \\
  \Gamma[H\to WW(\gamma)]& = \Gamma_\textrm{LO}[H\to WW]\left( 1 +\Delta_{HWW}^\textrm{EW} - \Delta r +\Delta_\textrm{brem}^{} 
  \right),  
  \label{eq:H to VV}
\end{align}
where $\Gamma_\textrm{LO}[H \to VV]$ is calculated as 
\begin{align}
 \Gamma_\textrm{LO}^{}[H\to VV] &=\left(\Gamma_{HVV}^\textrm{Tree, 1}\right)^2 \frac{m_H^3}{64\pi c_V^{}m_V^4}\left(1-4\frac{m_V^2}{m_H^2}
 +12\frac{m_V^4}{m_H^4}\right)\sqrt{1-4\frac{m_V^2}{m_H^2}}, 
\end{align}
with $c_V^{}=1$ ($2$) for $W$ ($Z$). 
In Eq.~(\ref{eq:H to VV}), $\Delta_{HVV}^\textrm{EW}$ indicates EW loop contributions expressed as 
\begin{align}
  \Delta_{HVV}^\textrm{EW}&=
 \frac{2}{\Gamma_{HVV}^{\textrm{Tree}, 1}}\textrm{Re}\left[\tilde{\Gamma}_{HVV}^\textrm{\textrm{1-loop}, 1} \right] \notag\\
 &+ \frac{1}{\Gamma_{HVV}^{\textrm{Tree}, 1}}\textrm{Re}\left[\Gamma_{HVV}^{\textrm{1-loop}, 2}\right]\frac{m_H^2}{m_V^2}\left(1-6\frac{m_V^2}{m_H^2}+8\frac{m_V^4}{m_H^4}\right)
     \left(1-4\frac{m_V^2}{m_H^2}+12\frac{m_V^4}{m_H^4}\right)^{-1}, \label{eq:Delta_HVV_EW}
\end{align}
with
\begin{align}
   \tilde{\Gamma}_{HVV}^{\textrm{1-loop}, 1} & = \Gamma_{HVV}^{\textrm{1-loop}, 1} -\Gamma_{HVV}^{\textrm{Tree}, 1}\frac{d}{dp^2}\hat{\Pi}_{VV}[p^2]\big|_{p^2=m_V^2}^{}. 
\end{align}
In the above expression, 
the second term is the contribution from wave function renormalizations of external vector bosons,
which are non-zero in our on-shell renormalization scheme. 
The tree level contribution of the $HVV$ vertex $\Gamma_{HVV}^{\textrm{Tree}, 1}$ is given in Eq.~(\ref{eq:HVV_tree}), and the explicit formula of $\hat{\Pi}_{VV}^{}$ is given in Eq.~(56) of Ref.~\cite{Kanemura:2015mxa}. 
The term $\Delta_\textrm{brem}$ in Eq.~(\ref{eq:H to VV}) 
indicates the contribution from the real photon bremsstrahlung which is needed in order to remove infrared (IR) divergence  
from virtual photon loop diagrams.  
The explicit formula of the contribution is given by~\cite{Kniehl:1993ay}   
\begin{align}
  \Delta_\textrm{brem} = &\frac{\alpha_\textrm{em}^{}}{\pi}\Bigg\{
  \left(\frac{2 -1/r}{\sqrt{1-1/r}}\log\rho_+ -1\right)\log\frac{m_W^2}{m_\gamma^2} -2\log(\rho_+ + \rho_- )
  -4\log(\rho_+ - \rho_- ) +\frac{14}{3} \notag\\
  &+\frac{2\log\rho_+}{\sqrt{1-1/r}} + \frac{2-1/r}{\sqrt{1-1/r}}
  \Bigg[\textrm{Li}_2(\rho_-^2) +\textrm{Li}_2(\rho_-^4) -2\zeta(2) -\frac{2\log\rho_+}{4r^2 -4r+3}\notag\\
  &+\log\rho_+ \left(5\log\rho_+ -2\log(\rho_+ +\rho_-) -2\log(\rho_+ -\rho_-)\right) \Bigg]\Bigg\},
  \label{eq:soft-photon}
\end{align}
with $r=m_H^2/(4m_W^2)$ and $\rho_\pm =\sqrt{r}\pm \sqrt{r-1}$.
In Eq.~(\ref{eq:soft-photon}), $m_\gamma$ represents the mass of the photon as a regulator. 
We numerically check that the $m_\gamma$ dependence is canceled by the virtual photon loop contributions in Eq.~(\ref{eq:Delta_HVV_EW}).

\section{Numerical results}\label{sec:Results}
In this section, we discuss numerical results of the decay BRs of $H$.
For $H\to hh$, $H\to f\bar{f}$ and $H\to VV$, we implement the EW and scalar corrections discussed in the previous section,
while the decay rates for the loop induced processes are calculated at LO in EW.
We take into account the QCD-NNLO corrections to the decay rate of $H\to q\bar{q}$ and loop induced processes. 
We impose constraints from perturbative unitarity, vacuum stability and data of the $S$ and $T$ oblique parameters in the following numerical calculations. 
The renormalization scale for the renormalized triple scalar vertices is set to be $\mu =m_H^{}$. 
In particular, we investigate the radiative corrections to the $H\to hh$ decay in detail. 
As discussed in Sec.~\ref{sec:Decay_rate},
the size of radiative corrections to the $H\to hh$ decay is dominantly determined by the top quark loop and non-decoupling effects of the additional Higgs bosons, so that it does not depend on the types of Yukawa interaction. 
Thus, we numerically evaluate the decay BRs with radiative corrections focusing on the Type-I THDM. 

\begin{figure}[t]
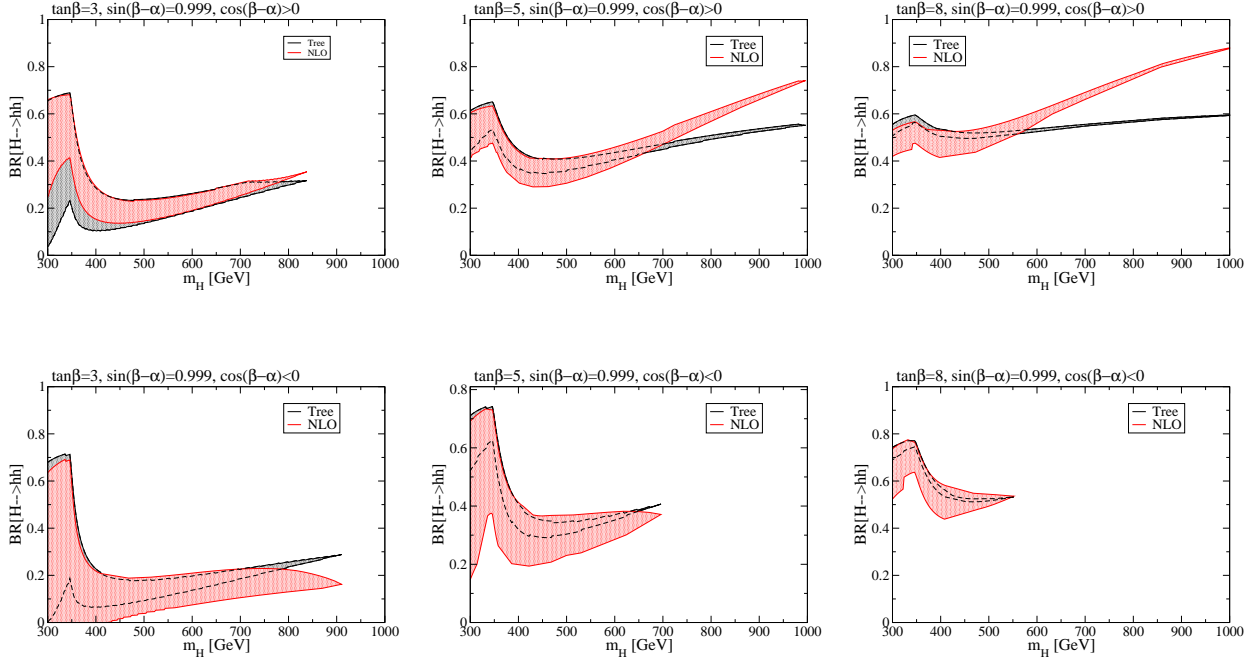

 \centering
 \includegraphics[width=52mm]{mbHdep_tan3_999_sei.eps}\hspace{0.3cm}
  \includegraphics[width=52mm]{mbHdep_tan5_999_sei.eps}\hspace{0.3cm}
   \includegraphics[width=52mm]{mbHdep_tan8_999_sei.eps}\vspace{1.0cm}
  \includegraphics[width=52mm]{mbHdep_tan3_999_hu.eps}\hspace{0.3cm}
  \includegraphics[width=52mm]{mbHdep_tan5_999_hu.eps}\hspace{0.3cm}
 \includegraphics[width=52mm]{mbHdep_tan8_999_hu.eps}
 \caption{BR of the $H\to hh$ decay with $\sin(\beta-\alpha)=0.999$ at NLO (red regions) and LO (black regions) in each value of
   $m_H^{} (=m_A^{}=m_{H^\pm}^{})$ in the Type-I THDM. 
   From left to right panels, the value of $\tan\beta$ are fixed to be 3, 5, and 8. 
   Upper panels and bottom panels show predictions with $\cos(\beta-\alpha) > 0$ and $\cos(\beta-\alpha) < 0$, respectively. 
   We scan the value of $M^2$ in the range of $M^2 \geq 0$ under the constraints of perturbative unitarity, vacuum stability and the $S$ and $T$ parameters. }
  \label{fig:nume1}
\end{figure}

\subsection{Branching ratio of $H\to hh$}\label{sec:BR of H to hh}
We first investigate the BR of the $H\to hh$ process. 
In order to see the structure of the loop corrections, we further simplify the approximate formula given in Eq.~(\ref{eq:Delta_0_B}) in Sec.~\ref{sec:Decay_rate} by taking degenerate masses of the additional Higgs bosons. 
In addition,  for $m_H^2 \gg m_h^2$, the bosonic loop contributions can be expanded as  
\begin{align}
  \Delta_{0, B}^{\textrm{EW}} &=
  \frac{\cot 2\beta}{16\pi^2}\frac{m_H^4}{v^4}\left(1-\frac{M^2}{m_H^2}\right)^2\Bigg\{
    24(1-C)\left(1 -\frac{M^2}{m_H^2}\right)+6C\notag\\
    & +\epsilon\left[8(2-3C)\left(1-\frac{M^2}{m_H^2}\right) +23-42C+3C\frac{m_H^2}{m_H^2 -M^2}\right]
    + \mathcal{O}(\epsilon^2)
    \Bigg\}, 
  \label{eq:Delta_0_B_2}
\end{align}
with $\epsilon\equiv m_h^2/m_H^2$ and $C\equiv 2 -\pi/\sqrt{3}\simeq 0.186$. From this approximate formula,
it is seen that $\Delta_{0, B}^{\textrm{EW}}$ is enhanced by $m_H^4$ for the case with $M\ll m_H$ due to the non-decoupling effect.
On the other hand, for $M\simeq m_H^{}$ such an enhancement is highly suppressed by the factor of $(1-M^2/m_H^2)^2$ and thus $\Delta_{0, B}^{\textrm{EW}}$ is roughly proportional to $(m_H^2 - M^2)^2/v^4$. 
It can also be seen that Eq.~(\ref{eq:Delta_0_B_2}) has the factor $\cot 2\beta$, 
so that the magnitude of the NLO corrections grows as $\tan\beta$ increases. 
We note that the NLO contributions come from the cross term of the amplitude from the tree level and one-loop contributions, 
so that the sign of the NLO contributions changes depending on the sign of $c_{\beta-\alpha}^{}$.
Namely, if $c_{\beta-\alpha}^{}$ is positive (negative), the NLO contributions increase (decrease) the decay rate.
Features of the loop corrections as those described here can be concretely confirmed by the following Figs.~\ref{fig:nume1} and \ref{fig:nume2}. 

\begin{figure}[t]
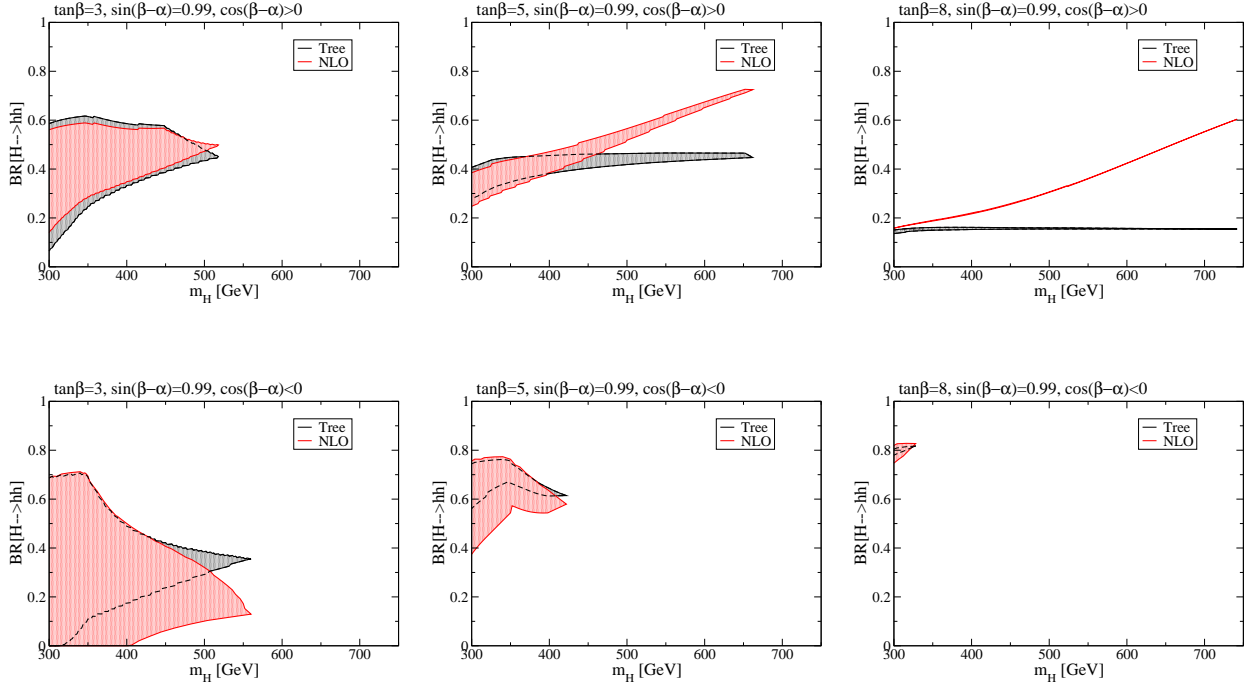

 \centering
  \includegraphics[width=52mm]{mbHdep_tan3_99_sei.eps}\hspace{0.3cm}
  \includegraphics[width=52mm]{mbHdep_tan5_99_sei.eps}\hspace{0.3cm}
  \includegraphics[width=52mm]{mbHdep_tan8_99_sei.eps}\vspace{1.0cm}
  \includegraphics[width=52mm]{mbHdep_tan3_99_hu.eps}\hspace{0.3cm}
  \includegraphics[width=52mm]{mbHdep_tan5_99_hu.eps}\hspace{0.3cm}
  \includegraphics[width=52mm]{mbHdep_tan8_99_hu.eps}
 \caption{Same as Fig.~\ref{fig:nume1}, but for $\sin(\beta-\alpha)^{}=0.99$.  }
  \label{fig:nume2}
\end{figure}

Fig.~\ref{fig:nume1} shows the $m_H$ dependence of the BR of $H\to hh$ including the NLO corrections (red regions) and that at tree level (black regions) for the degenerate mass case; i.e., $m_H=m_A=m_{H^\pm}$.      
We fix $s_{\beta-\alpha}^{}=0.999$, where the upper panels (the lower panels) represent results with $c_{\beta-\alpha}^{}>0$ ($c_{\beta-\alpha}^{}<0$).
Results for $\tan\beta=3$, 5 and 8 are shown from the left panels to the right panels.
We scan the parameter $M^2$ within $M^2 \geq 0 $ in each value of $m_H^{}$.
In the case with $c_{\beta-\alpha}^{}> 0$, it can be confirmed that the NLO corrections typically increase the BR, which tends to be clearer at large $\tan\beta$ and/or large mass regions.  
For $c_{\beta-\alpha}^{} < 0$, the NLO corrections typically decrease the BR.
We note that the parameter regions where the non-decoupling quantum effects are important are excluded by the perturbative unitarity bound. 
The value of the BR drops sharply at around $m_H = 350$ GeV because the $H\to t\bar{t}$ process opens.

Fig.~\ref{fig:nume2} shows the decay BR of $H\to hh$ in the case with $s_{\beta-\alpha}^{}=0.99$, while the other configurations are the same as those in Fig.~\ref{fig:nume1}. 
As compared with the case for $s_{\beta-\alpha}^{}=0.999$, the BR typically becomes smaller values and the upper limit on $m_H^{}$ is stronger due to the theoretical constraints.

\begin{figure}
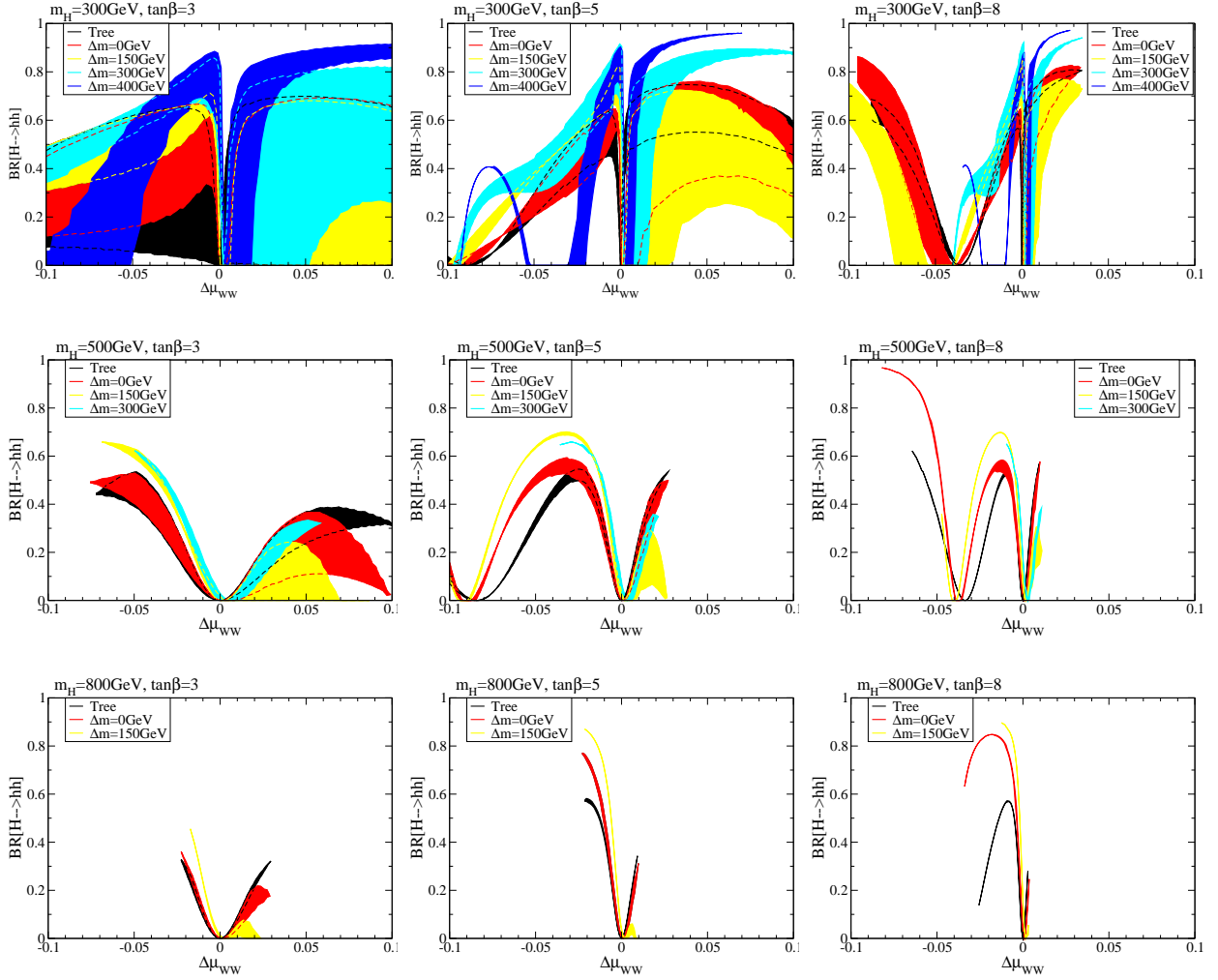

 \centering
 \includegraphics[width=54mm]{hWW_BR_mbH300_tan3.eps}
 \includegraphics[width=54mm]{hWW_BR_mbH300_tan5.eps}
 \includegraphics[width=54mm]{hWW_BR_mbH300_tan8.eps}\vspace{0.5cm}
 \includegraphics[width=54mm]{hWW_BR_mbH500_tan3.eps}
 \includegraphics[width=54mm]{hWW_BR_mbH500_tan5.eps}
 \includegraphics[width=54mm]{hWW_BR_mbH500_tan8.eps}\vspace{0.5cm}
 \includegraphics[width=54mm]{hWW_BR_mbH800_tan3.eps}
 \includegraphics[width=54mm]{hWW_BR_mbH800_tan5.eps}
 \includegraphics[width=54mm]{hWW_BR_mbH800_tan8.eps} 
 \caption{Correlation between the BR of $H\to hh$ and the deviation in the ratio of the branching ratio of $h \to WW^*$ from the SM prediction in the Type-I THDM.
   Black regions are predictions at tree level, while red, yellow, cyan and blue regions express predictions with the NLO corrections in the case with
   $\Delta m~ (\equiv m_A-m_H) =$ 0, 150 GeV, 300 GeV and 400 GeV, respectively. 
   Top, middle and bottom panels show results where $m_H$ is fixed to be 300 GeV, 500 GeV and 800 GeV, respectively. 
   From the panel of the left, the value of $\tan\beta$ is fixed to be 3, 5, and 8. 
   We scan the value of $M^2$ and $\cos(\beta-\alpha)$ under the constraints of perturbative unitarity, vacuum stability and the $S$ and $T$ parameters. }
  \label{fig:sss300_500_800}
\end{figure}

Next, we investigate the correlation between the BR of $H\to hh$ and the BR of $h\to WW^*\to Wf\bar{f}'$ 
which is expected to be measured with 2\% accuracy at the ILC with the collision energy being 250 GeV~\cite{Fujii:2017vwa}.  
In order to parametrize the deviation in the BR of $h\to WW^*$ from the SM prediction,  
we introduce $\Delta\mu_{WW}^{}$ defined as 
\begin{align}
  \Delta\mu_{WW}^{} \equiv \frac{\textrm{BR}[h\to WW^*]_\textrm{THDM}}{\textrm{BR}[h\to WW^*]_\textrm{SM}}-1,
  \label{eq:delta_mu_ww}
\end{align}
where $\textrm{BR}[h\to WW^*]_\textrm{THDM}$ ($\textrm{BR}[h\to WW^*]_\textrm{SM}$) represents the BR in the THDM (SM). 
We numerically evaluate the value of $\Delta\mu_{WW}^{}$ by using \texttt{H-COUP}~\cite{Kanemura:2019slf}.

In Fig.~\ref{fig:sss300_500_800}, we show the correlation between the BR of the $H\to hh$ process and $\Delta\mu_{WW}^{}$ for each fixed value of $\tan\beta$ and $m_H^{}$. 
The values of $s_{\beta-\alpha}^{}$ and $M^2$ are scanned under the constraints of perturbative unitarity, vacuum stability and data of the $S$ and $T$ oblique parameters.
The regions shaded in black show the results at LO, while those shaded in red, yellow, cyan and blue show the results at NLO with the mass difference $\Delta m= m_A-m_H$ to be 0, 150, 300 and 400 GeV, respectively.
In some of the panels, several colored regions do not appear because of no allowed region by the constraints.  
Results with $\Delta\mu_{WW}^{}<0$ ($\Delta\mu_{WW}^{}>0$) correspond to those with $c_{\beta-\alpha}^{}>0$ ($c_{\beta-\alpha}^{}<0$), 
because the partial decay width of $h\to b\bar{b}$ is enhanced (suppressed). 
In the case with heavier $H$, predictions are well determined to be narrower regions, because the allowed range of $M^2$ by the theoretical constraints is shrunk. 
It is seen that for $\Delta m=0$ and $\Delta\mu_{WW}^{} <0$ ($\Delta\mu_{WW}^{} >0$) , the value of the BR is pushed up (down) by the NLO corrections, 
in which this behavior can be understood from Eq.~(\ref{eq:Delta_0_B_2}) and is consistent with the results shown in Figs.~\ref{fig:nume1} and \ref{fig:nume2}. 
It can also be seen that if the value of $\tan\beta$ increases, the NLO corrections increase, because bosonic-loop effects are proportional to $\cot2\beta$ in Eq.~(\ref{eq:Delta_0_B_2}). 
For $\Delta\mu_{WW}^{} <0$; i.e., $c_{\beta-\alpha}^{}> 0$, the BR becomes 0 at particular values of $\Delta\mu_{WW}^{}$; e.g., at around $\Delta\mu_{WW}^{} = -0.4$ for $m_H = 300$ GeV and $\tan\beta = 8$.
Such behavior can be explained by the expression of the tree level $Hhh$ coupling $\lambda_{Hhh}^{}$ as given in Eqs.~(\ref{eq:ppp_series}) and (\ref{eq:Hhh_series}).

In the case with non-zero  mass difference among additional Higgs bosons; i.e., $\Delta m\neq 0$,
the behavior of loop corrections can drastically be different from that in case with $\Delta m=0$. 
As $\Delta m$ increases, the difference from the tree level prediction becomes more significant than that in the degenerate mass case. 
For results of $m_H=300$ GeV, predictions including the NLO corrections can be about 30 \% larger than tree level predictions
if $\Delta m$ is larger than 300 GeV. 
If $\Delta m$ is non-zero, the effect of NLO corrections can increase the BR in the both cases with $c_{\beta-\alpha}^{}>0$ and $c_{\beta-\alpha}^{}<0$.
However, due to theoretical constraints, allowed regions become smaller as $m_H$ increases.

If $\Delta\mu_{WW}^{}$ is larger than 2\%, it can be observed as a deviation from the SM prediction by the precision measurements at the ILC with the center of mass energy $\sqrt{s}$ to be 250 GeV~\cite{Fujii:2017vwa}.  
However, even if the deviation of the $h\to WW^*$ decay is too small to be observed by the ILC, it might be possible to explore $H$ via the $H\to hh$ process at the HL-LHC. 
If $H$ is lighter than $2m_t$ and $\Delta m$ is non-zero, the $H\to hh$ decay mode can be dominant in large parameter regions. 

\begin{figure}
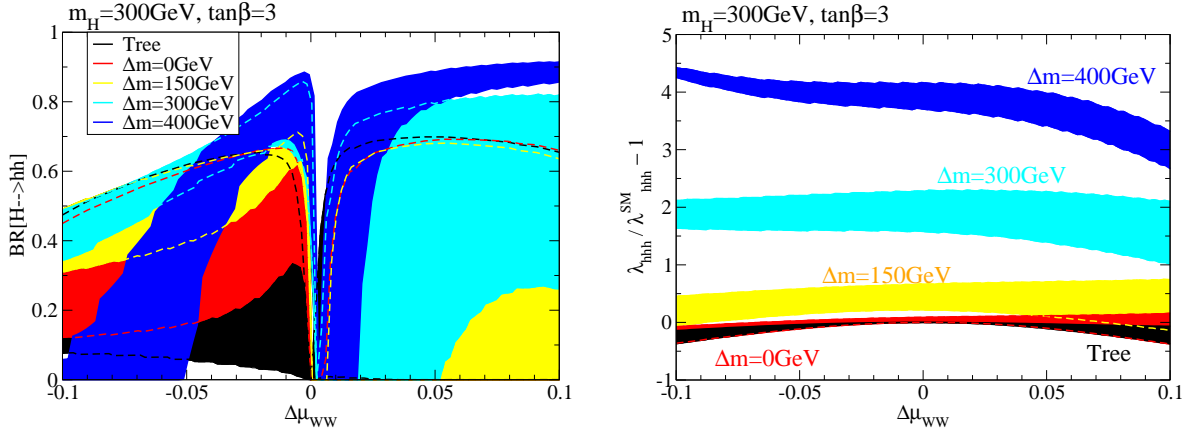

 \centering
  \includegraphics[width=75mm]{hWW_BR_mbH300_tan3.eps}\hspace{0.5cm}
 \includegraphics[width=75mm]{hWW_hhh_mbH300_tan3.eps}
 \caption{Correlation between BR($H\to hh$) and $\Delta\mu_{WW}^{}$ (left panel) and that between BR($H\to hh$) and $\lambda_{hhh}^{}/\lambda_{hhh}^\textrm{SM} -1$ (right panel) for $m_H=300$ GeV and $\tan\beta=3$. The color definitions of the regions are the same as specified in Fig.~\ref{fig:sss300_500_800}. We scan the value of $M^2$ and $\cos(\beta-\alpha)$ under the constraints of perturbative unitarity, vacuum stability and the $S$ and $T$ parameters. For the $hhh$ vertex, the renormalization scale is also taken as $\mu=m_H^{}$ in the same as that for $\Gamma[H\to hh]$. }
  \label{fig:hWW_hhh}
\end{figure}

It is known that similar non-decoupling effects also appear in the loop corrected $hhh$ coupling~\cite{Kanemura:2004mg,Kanemura:2015mxa,Kanemura:2017wtm}. 
The physics of the $hhh$ coupling is strongly related with the EW baryogenesis, because the strong first order phase transition can lead to 
a large deviation in the $hhh$ coupling from the SM prediction at zero temperature~\cite{Kanemura:2002vm,Kanemura:2004ch,Grojean:2004xa,Braathen:2019zoh,Braathen:2020vwo}.
The $hhh$ coupling can be extracted from the measurements of the double-Higgs production at hadron, lepton and photon colliders as discussed in Ref.~\cite{Asakawa:2010xj}. 
The measurement accuracy of the $hhh$ coupling is expected to be about 27\% at the ILC with $\sqrt{s}=500$ GeV~\cite{Fujii:2017vwa}. 
In Fig.~\ref{fig:hWW_hhh}, we show the correlation between $\Delta\mu_{WW}^{}$ and the deviation in the renormalized $hhh$ vertex in the Type-I THDM from that in the SM, in the case with $m_H^{}=300$ GeV and $\tan\beta=3$. 
We calculate the renormalized $hhh$ vertex using \texttt{H-COUP}~\cite{Kanemura:2017gbi,Kanemura:2019slf},
excepting parameter regions causing BR($H\to hh$)$<0$.   
The color definitions of the regions are the same as specified in Fig.~\ref{fig:sss300_500_800}.
In order to examine the correlation with the BR of $H\to hh$, we also place the panel which is shown in Fig.~\ref{fig:sss300_500_800}.
It can be seen that the deviation of the $hhh$ coupling is almost determined by the magnitude of $\Delta m$. 
The larger $\Delta m$ causes the larger deviation in the $hhh$ coupling, since it is caused by a larger non-decoupling effect.
Namely, the structure of the non-decoupling effects is the same as those of the $H\to hh$ decay. 
Such parameter regions are common with regions where the $hhh$ coupling shifts from the SM predictions significantly so that the $H\to hh$ search at the HL-LHC might also be used to test the EW baryogenesis scenario multi-directionally.

\subsection{Branching ratios of $H$}\label{sec:BR of H}
\begin{figure}
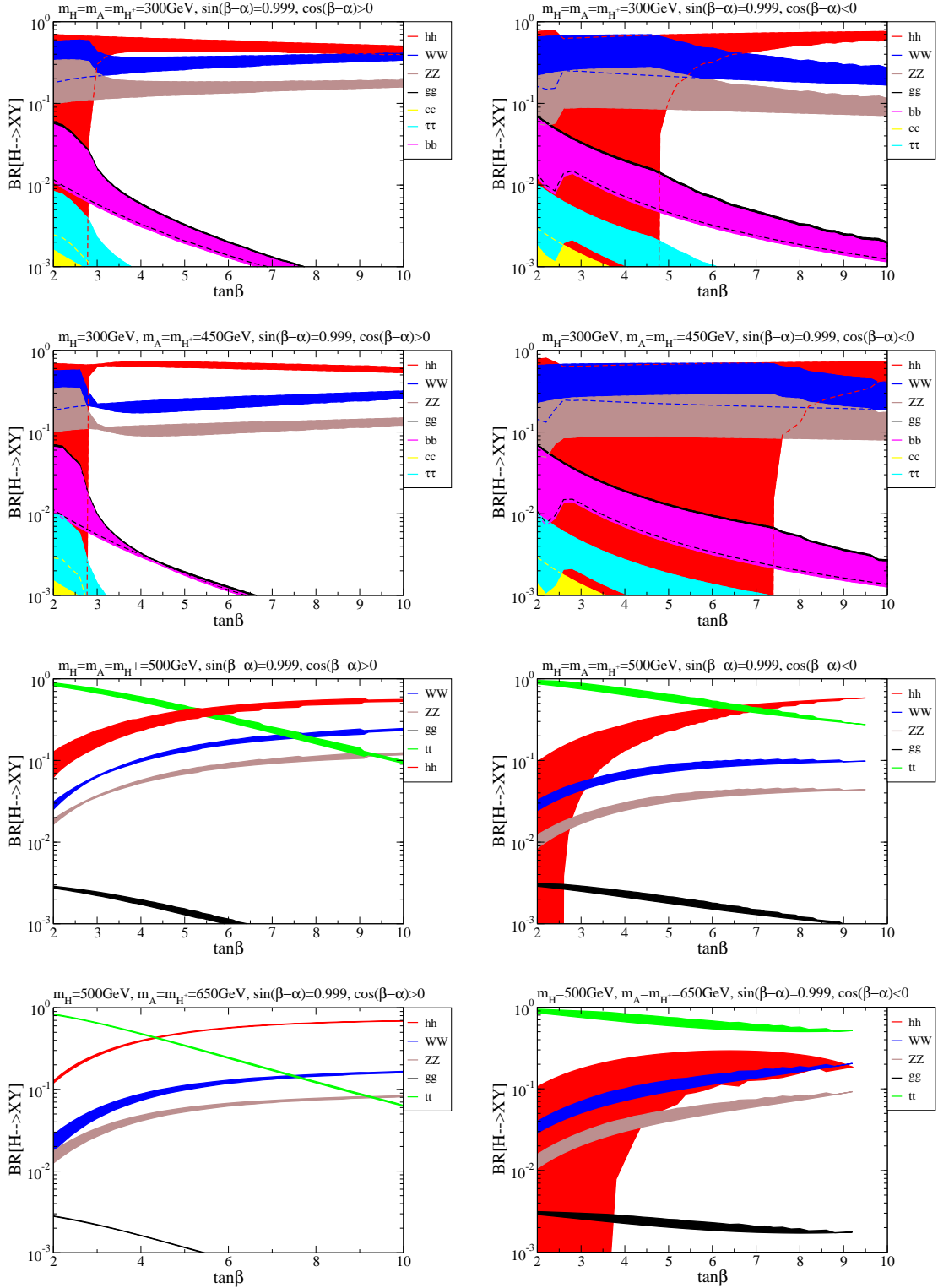

 \centering
 \includegraphics[width=70mm]{tandep_300_999_dm0_sei.eps}\hspace{0.5cm}
 \includegraphics[width=70mm]{tandep_300_999_dm0_hu.eps}\vspace{0.4cm}
 \includegraphics[width=70mm]{tandep_300_999_dm150_sei.eps}\hspace{0.5cm}
 \includegraphics[width=70mm]{tandep_300_999_dm150_hu.eps}\vspace{0.4cm}
  \includegraphics[width=70mm]{tandep_500_999_dm0_sei.eps}\hspace{0.5cm}
  \includegraphics[width=70mm]{tandep_500_999_dm0_hu.eps}\vspace{0.4cm}
 \includegraphics[width=70mm]{tandep_500_999_dm150_sei.eps}\hspace{0.5cm}
 \includegraphics[width=70mm]{tandep_500_999_dm150_hu.eps}\vspace{0.4cm}  
 \caption{Branching ratios for $H$ as a function of $\tan\beta$ in the case of $\sin(\beta-\alpha)=0.999$ with $\cos(\beta-\alpha)>0$ (left panels) and $\cos(\beta-\alpha)<0$ (right panels). 
The upper four and lower four panels show the case with $m_H=300$ GeV and 500 GeV, respectively. The first and third (second and fourth) panels from the top show the case with $\Delta m$ ($=m_A^{} - m_{H^\pm}^{}$) $= 0$ (150 GeV). The value of $M^2$ is scanned under the constraints of perturbative unitarity, vacuum stability and the $S$ and $T$ parameters. }
  \label{fig:BR_H}
\end{figure}

Finally, we investigate the other decay modes of $H$ in the Type-I THDM. 
In Fig.~\ref{fig:BR_H}, we show the decay BRs for $H$ as a function of $\tan\beta$ for $s_{\beta-\alpha}^{}=0.999$ and $c_{\beta-\alpha}^{}>0$ ($c_{\beta-\alpha}^{}<0$) in the left (right) panels.
We take $(m_H, \Delta m)=(300, 0)$ GeV, $(300, 150)$ GeV, $(500,0)$ GeV and $(500,150)$ GeV from the top to the bottom panels.  
The value of $M^2$ is scanned with $M^2\geq 0$ in all the panels. 
As we can see in the panels for $m_H=300$ GeV, $H\to VV$ are the dominant decay modes.
The BR of $H\to hh$ can also be dominant depending on the values of $M^2$ and $\tan\beta$.
In the low $\tan\beta$ regions, the wider range of $M^2$ is allowed by the theoretical constraints, so that possible values of BR($H\to hh$) spread.
For $\Delta m =150$ GeV and $c_{\beta-\alpha}>0$ ($c_{\beta-\alpha}<0$), the loop effects enhance (suppress) the decay rate of the $H\to hh$ process as compared with the case for $\Delta m=0$, so that BR($H\to hh$) tends to be more important than BR($H\to VV$). 
In the case with $m_H^{}=500$ GeV, the $H\to t\bar{t}$ mode opens, whose decay rate is proportional to $\cot^2\beta$. 
Thus, the $H\to hh$ process becomes the main decay mode in the large $\tan\beta$ region. 
For $\Delta m =150$ GeV and $c_{\beta-\alpha}>0$, BR($H\to hh$) is typically enhanced with several tens of percent than that for $\Delta m=0$. 
On the other hand for $c_{\beta-\alpha}^{}<0$, BR($H\to hh$) does not increase even if $\tan\beta$ becomes large and/or $\Delta m$ is taken to be non-zero. 
Therefore, $H\to t\bar{t}$ is typically the main decay mode. 
We note that the search for heavy Higgs bosons decaying into $t\bar{t}$ at hadron colliders is challenging due to the large the SM background,
but various simulation studies for detecting such Higgs bosons have been done at LHC in Refs.~\cite{Craig:2015jba,Bernreuther:2015fts,Kanemura:2015nza,Djouadi:2016ack,Hespel:2016qaf,Carena:2016npr,Bernreuther:2017yhg,BuarqueFranzosi:2017jrj,Adhikary:2018ise,Djouadi:2019cbm,Bahl:2020kwe}.

\section{Discussions}\label{sec:Discussions}
We discuss the direct search for the additional Higgs bosons at future collider experiments.
At the HL-LHC, $H$ or $A$ is mainly produced by the gluon fusion process and the associated production with $b\bar{b}$.~\footnote{The pair productions of the additional Higgs bosons such as $pp\to HA/H^\pm H/H^\pm A/ H^+H^-$ can also be important for the direct searches at the HL-LHC, whose cross sections are simply determined by their masses, see e.g., Ref.~\cite{Kanemura:2021dez}.}
The parameter region expected to be explored via these single productions has been studied in Ref.~\cite{Aiko:2020ksl},  
where the analysis has been done at LO in the EW interaction.
It goes without saying that the search for the additional Higgs bosons can also be done at the ILC energy upgrade, where the collision energy $\sqrt{s}$ can be extended to be up to 1 TeV~\cite{Baer:2013cma}. 
They can mainly be produced in pairs as $e^+ e^- \to HA$ and $e^+ e^- \to H^+H^-$ up to 500 GeV for the degenerate mass case.  
%
As we have shown in this paper, the BRs of $H$ can significantly be changed by the NLO corrections in EW and scalar interactions. 
Thus, it is quite important to include such effects in the exploration of the additional Higgs bosons at the HL-LHC and the ILC. 
We will upgrade the numerical program \texttt{H-COUP}~\cite{Kanemura:2017gbi,Kanemura:2019slf} such that the decay BRs for the additional Higgs bosons are calculated including EW, scalar and QCD corrections based on this paper ($H$), Ref.~\cite{Aiko:2022XX} ($A$) and Ref.~\cite{Aiko:2021can} ($H^\pm$).
We will then be able to discuss the synergy between the direct search and the precise measurement of $h$, which will be left for future works.


Finally, we would like to comment that a portion of the parameter space shown in our numerical results is excluded by the current experimental data from the additional Higgs boson searches at LHC and the measurement of the signal strength of $h$ given in Sec.~\ref{sec:Model}.  
For example, the region $\Delta\mu_{WW}^{}\lesssim 0$ are excluded by taking into account the constraints on $\alpha$ and $\beta$ in the THDMs from the signal strength data~\cite{ATLAS:2021vrm}.
It is, however, seen that there is a large discrepancy between the region excluded by the observed data and that by the expectation of the MonteCarlo analysis. 
Thus, the observed exclusion can be drastically changed by accumulating more data. 
Therefore, we have investigated wider regions of the parameter space than the allowed ones by the current experimental data in the numerical calculations.

\section{Conclusions}\label{sec:Conclusions}
We have computed the decay rates of the additional CP-even Higgs boson $H$; i.e., $H\to hh$, $H\to f\bar{f}$ and $H\to VV$ with the EW and scalar NLO corrections in the THDMs with a softly broken $Z_2$ symmetry, where QCD corrections are also included for $H\to q\bar{q}$.   
For loop induced processes, we have calculated the decay rates at LO in EW and scalar interactions, but including QCD corrections.
We have particularly focused on the scenario with the nearly alignment in the Type-I THDM for numerical evaluations.   
We have clarified that various parameter dependences such as $\tan\beta$, $m_H$, $\Delta m (=m_A-m_H)$ and the sign of $c_{\beta-\alpha}^{}$ on the BR of $H\to hh$ under the constraints from perturbative unitarity, vacuum stability and electroweak precision data. 
It has been found that the effect of the radiative corrections on the BR of $H\to hh$ can drastically change its LO prediction due to the non-decoupling effect of the additional Higgs boson loops. 
We have also investigated the correlation between the deviation in the BR of $h\to WW^*$ from the SM prediction ($\Delta\mu_{WW}^{}$) and the BR of $H\to hh$ at NLO in EW and scalar interactions. 
For example, in the case with $m_H=500$ GeV and $\tan\beta=3$ and $\Delta\mu_{WW}^{}=0.05$, BR($H\to hh$) can be 0.3-0.4, 0.1-0.4, 0-0.25 and 0.3-0.35 at LO, at NLO with $\Delta m=0$, 150 GeV and 300 GeV, respectively. 
Even if $|\Delta\mu_{WW}^{}|$ is less than 0.02 which might not be able to be detected at the ILC, we have seen that the $H\to hh$ can be the dominant decay mode; e.g., BR($H\to hh$) can be about 70 \% for $m_H=500$ GeV, $\Delta m=300$ GeV and $\tan\beta=5$. 
Therefore, it has been shown that including the radiative corrections to the decay of $H$ is quite important for the direct searches for the additional Higgs bosons at future collider experiments such as the HL-LHC and the ILC.

\section*{Acknowledgment} 
This work is supported in part by the Grant-in-Aid on Innovative Areas, the Ministry of Education, Culture, Sports, Science and Technology, No.~16H06492, and by the JSPS KAKENHI Grant No.~20H00160 [S.K.],  
Early-Career Scientists, No.~20K14474 [M.K.] and
Early-Career Scientists, No.~19K14714 [K.Y.]. 

\appendix
\section{One-loop diagrams for the renormalized $Hhh$ vertex} 
We present analytic formulae of one-loop diagrams related with the renormalized $Hhh$ vertex.
All feynman diagrams are computed in the 't Hooft-Feynman gauge, and are expressed by Passarino Veltman functions~\cite{Passarino:1978jh}.

Two-point functions include not only 1PI contributions $\Pi_{h_i h_j}^\textrm{1PI}$ but also pinch terms $\Pi_{h_i h_j}^\textrm{PT}$ and tadpole contributions $\Pi_{h_i h_j}^\textrm{Tad}$. 
For two-point functions of scalar fields, explicit formulae of $\Pi_{h_i h_j}^\textrm{1PI}$ and those of $\Pi_{h_i h_j}^\textrm{PT}$ 
are given in Ref.~\cite{Kanemura:2015mxa} and in Ref.~\cite{Kanemura:2017wtm}, respectively. 
The contributions $\Pi_{h_i h_j}^\textrm{Tad}$ are calculated as  
\begin{align}
 &\Pi_{hh}^\textrm{Tad} = \frac{6\lambda_{hhh}}{m_h^2}T_h^\textrm{1PI} + \frac{2\lambda_{Hhh}}{m_H^2}T_H^\textrm{1PI}, \\
 &\Pi_{HH}^\textrm{Tad} = \frac{2\lambda_{HHh}}{m_h^2}T_h^\textrm{1PI} + \frac{6\lambda_{HHH}}{m_H^2}T_H^\textrm{1PI}, \\
 &\Pi_{Hh}^\textrm{Tad} = \frac{2\lambda_{Hhh}}{m_h^2}T_h^\textrm{1PI} + \frac{2\lambda_{HHh}}{m_H^2}T_H^\textrm{1PI}, \\
  &\Pi_{AG}^\textrm{Tad} = \frac{\lambda_{hGA}}{m_h^2}T_h^\textrm{1PI} + \frac{\lambda_{HGA}}{m_H^2}T_H^\textrm{1PI},  
\end{align}
where explicit formulae of $T_{h_i}^\textrm{1PI}$ are given in Ref.~\cite{Kanemura:2015mxa}. 

Contributions of one-loop diagrams for the three-point vertex $Hhh$ are also composed by 1PI diagram contributions and tadpole contributions as 
\begin{align}
  \Gamma_{Hhh}[p_1^2,p_2^2,q^2] = \Gamma_{Hhh}^\textrm{1PI}[p_1^2,p_2^2,q^2] + \Gamma_{Hhh}^\textrm{Tad}, 
\end{align}
where $p_1^\mu$ and $p_2^\mu$ are the incoming momenta of two 125 GeV Higgs bosons $h$, and 
$q^\mu (=p_1^\mu + p_2^\mu)$ is the outgoing momentum of $H$. 
Tadpole contributions to the $Hhh$ vertex are given as 
\begin{align} 
 &\Gamma_{Hhh}^\textrm{Tad}= 6\frac{\lambda_{Hhhh}^{}}{m_h^2}T_h^\textrm{1PI} + 4\frac{\lambda_{HHhh}^{}}{m_H^2}T_H^\textrm{1PI}. 
\end{align} 
Fermionic loop contributions and bosonic loop contributions for the 1PI diagrams are respectively calculated as 
\begin{align}
 &\Gamma_{Hhh}^\textrm{1PI}[p_1^2,p_2^2,q^2]_F^{}=
 -\sum_f^{}\frac{8m_f^4N_c^f}{16\pi^2v^3}(\kappa_{f}^f)^2\kappa_{f}^H
 \big[B_0^{}[p_1^2;f,f] + B_0^{}[p_2^2;f,f] +B_0^{}[q^2;f,f] \notag\\
   & +(4m_f^2 - q^2 +p_1\cdot p_2)C_0[f, f, f] \big], 
\end{align}
and 
 \begin{align}
 &(16\pi^2)\Gamma_{Hhh}^\textrm{1PI}[p_1^2,p_2^2,q^2]_B^{} \notag\\[8pt] &=
  \frac{g_Z^3 m_Z^{}}{4} \Big( s_{\beta - \alpha}^{2} c_{\beta - \alpha}^{} C^{\phi\phi\phi}_{VVS}[Z, Z, A] 
   - c_{\beta -\alpha}^3C^{\phi\phi\phi}_{VSV}[Z, A, Z] + s_{\beta - \alpha}^2 c_{\beta-\alpha}^{}C_{SVV}^{\phi\phi\phi}[A, Z, Z]\notag\\[8pt]
   &- s_{\beta -\alpha}^2 c_{\beta -\alpha}^{}C_{VVS}^{\phi\phi\phi}[Z , Z, G^0] 
   - s_{\beta -\alpha}^2 c_{\beta -\alpha}^{}C_{VSV}^{\phi\phi\phi}[Z, G^0, Z]
    - s_{\beta -\alpha}^2 c_{\beta -\alpha}^{}C_{SVV}^{\phi\phi\phi}[G^0, Z, Z] \Big)\notag\\[8pt]
  &+ \frac{g^3 m_W^{}}{2}\Big( s_{\beta -\alpha}^2 c_{\beta -\alpha}^{}C_{VVS}^{\phi\phi\phi}[W, W, H^\pm] 
 - c_{\beta-\alpha}^3 C_{VSV}^{\phi\phi\phi}[W, H^\pm, W] 
 + s_{\beta -\alpha}^2 c_{\beta -\alpha}^{}C_{SVV}^{\phi\phi\phi}[H^\pm, W, W]\notag\\[8pt]
  & - s_{\beta -\alpha}^2 c_{\beta -\alpha}^{}C_{VVS}^{\phi\phi\phi}[W, W, G^\pm] 
      - s_{\beta -\alpha}^2 c_{\beta -\alpha}^{}C_{VSV}^{\phi\phi\phi}[W, G^\pm, W] 
      - s_{\beta -\alpha}^2 c_{\beta -\alpha}^{}C_{SVV}^{\phi\phi\phi}[G^\pm, W, W]\Big)\notag\\[8pt]
  & + \frac{g^2}{2}\Big( s_{\beta-\alpha}^{}c_{\beta -\alpha}^{}\lambda_{hG^+G^-}^{}C_{VSS}^{\phi\phi\phi}[W, G^\pm, G^\pm]
      + s_{\beta-\alpha}^{2}\lambda_{HG^+G^-}^{}C_{SVS}^{\phi\phi\phi}[G^\pm, W, G^\pm]  \notag\\[8pt]
     &  + s_{\beta-\alpha}^{}c_{\beta -\alpha}^{}\lambda_{hG^+G^-}^{}C_{SSV}^{\phi\phi\phi}[G^\pm, G^\pm, W]
     -s_{\beta-\alpha}^{}c_{\beta -\alpha}^{}\lambda_{hH^+H^-}^{}C_{VSS}^{\phi\phi\phi}[W, H^\pm, H^\pm] \notag\\[8pt]
      &  + c_{\beta -\alpha}^{2}\lambda_{HH^+H^-}^{}C_{SVS}^{\phi\phi\phi}[H^\pm, W, H^\pm]
        - s_{\beta-\alpha}^{}c_{\beta -\alpha}^{}\lambda_{hH^+H^-}^{}C_{SSV}^{\phi\phi\phi}[H^\pm, H^\pm, W]\notag\\[8pt]
   &  + c_{\beta -\alpha}^{2}\lambda_{hG^+H^-}^{}C_{VSS}^{\phi\phi\phi}[W, H^\pm, G^\pm] 
        + s_{\beta-\alpha}^{}c_{\beta -\alpha}^{}\lambda_{HG^+H^-}^{}C_{SVS}^{\phi\phi\phi}[H^\pm,W, G^\pm]\notag\\[8pt]
    &  - s_{\beta-\alpha}^{2}\lambda_{hG^+H^-}^{}C_{SSV}^{\phi\phi\phi}[H^\pm, G^\pm, W]
         - s_{\beta-\alpha}^{2}\lambda_{hG^+H^-}^{}C_{VSS}^{\phi\phi\phi}[W, G^\pm, H^\pm] \notag\\[8pt]
     & + s_{\beta-\alpha}^{}c_{\beta -\alpha}^{}\lambda_{HG^+H^-}^{}C_{SVS}^{\phi\phi\phi}[G^\pm,W, H^\pm]
         + c_{\beta -\alpha}^{2}\lambda_{hG^+H^-}^{}C_{SSV}^{\phi\phi\phi}[G^\pm, H^\pm, W]\Big) \notag\\[8pt]
     & + \frac{g_Z^2}{2}\Big( - s_{\beta -\alpha}^{}c_{\beta- \alpha}^{}\lambda_{hAA}^{}C_{VSS}^{\phi\phi\phi}[Z, A, A]
         +  c_{\beta- \alpha}^2\lambda_{HAA}^{}C_{SVS}^{\phi\phi\phi}[A, Z, A] 
         - s_{\beta -\alpha}^{}c_{\beta- \alpha}^{}\lambda_{hAA}^{}C_{SSV}^{\phi\phi\phi}[A, A, Z]\notag\\[8pt]
     & + s_{\beta -\alpha}^{}c_{\beta- \alpha}^{}\lambda_{hG^0G^0}^{}C_{VSS}^{\phi\phi\phi}[Z, G^0, G^0]
         + s_{\beta -\alpha}^2\lambda_{HG^0G^0}^{}C_{SVS}^{\phi\phi\phi}[G^0, Z, G^0]
          +  s_{\beta -\alpha}^{}c_{\beta- \alpha}^{}\lambda_{hG^0G^0}^{}C_{SSV}^{\phi\phi\phi}[G^0, G^0, Z]\Big)\notag\end{align}
 \begin{align}
     & +\frac{g_Z^2}{4}\Big( c_{\beta- \alpha}^{2}\lambda_{hG^0 A}^{}C_{VSS}^{\phi\phi\phi}[Z, A, G^0] 
         +  s_{\beta -\alpha}^{}c_{\beta- \alpha}^{}\lambda_{HG^0A}^{}C_{SVS}^{\phi\phi\phi}[A, Z, G^0]
         - s_{\beta -\alpha}^2 \lambda_{hG^0 A}^{}C_{SSV}^{\phi\phi\phi}[A, G^0, Z] \notag\\[8pt]
     & - s_{\beta -\alpha}^{2}\lambda_{hG^0A}^{}C_{VSS}^{\phi\phi\phi}[Z, G^0, A]
          + s_{\beta -\alpha}^{}c_{\beta- \alpha}^{}\lambda_{HG^0A}^{}C_{SVS}^{\phi\phi\phi}[G^0, Z,  A]
          + c_{\beta- \alpha}^{2}\lambda_{hG^0A}^{}C_{SSV}^{\phi\phi\phi}[G^0, A, Z]\Big)\notag\\[8pt]
     & +\frac{g^3m_W^{3}}{2}s_{\beta -\alpha}^2 c_{\beta-\alpha}^{}\left(16C_0^{}[W,W,W] - C_0^{}[c^\pm,c^\pm,c^\pm] \right) 
        +\frac{g_Z^3m_Z^{3}}{4}s_{\beta -\alpha}^2 c_{\beta-\alpha}^{}\left(16C_0^{}[Z,Z,Z] - C_0^{}[c_Z,c_Z,c_Z] \right) \notag\\[8pt]
   & -72\lambda_{hhh}^2\lambda_{Hhh}^{}C_0[h,h,h]  
      -24 \lambda_{hhh}^{}\lambda_{Hhh}^{}\lambda_{HHh}(C_0[H, h,h] + C_0[h, h, H] )
      - 8 \lambda_{Hhh}^3 C_0[h, H, h]  \notag\\[8pt]
   & - 8 \lambda_{HHh}^2 \lambda_{Hhh}^{}(C_0[H, H, h] + C_0[h, H, H])  -24\lambda_{HHH}^{}\lambda_{Hhh}^2 C_0[H, h, H]
      - 24\lambda_{HHH}^{}\lambda_{HHh}^2 C_0[H, H, H] \notag \\[8pt]
   & - 8\lambda_{HAA}^{}\lambda_{hAA}^2 C_0[A, A, A]
     -2 \lambda_{HG^0A}^{}\lambda_{hG^0A}^{}\lambda_{hG^0G^0}^{} (C_0[A, G^0, G^0] + C_0[G^0, G^0, A]) \notag\\[8pt]
   & - 8 \lambda_{HG^0G^0}^{}\lambda_{hG^0G^0}^2 C_0[G^0, G^0, G^0] 
       -2 \lambda_{HG^0 G^0}^{}\lambda_{hG^0 A}^2 C_0[G^0, A, G^0] \notag\\[8pt]
   & -2 \lambda_{HG^0A}^{}\lambda_{hAA}^{}\lambda_{hG^0A}^{} (C_0[A,A,G^0] + C_0[G^0, A,  A] )
    -2\lambda_{HAA}^{}\lambda_{hG^0 A}^{2} C_0[A, G^0, A] \notag\\[8pt]
   & -2\lambda_{HG^+ G^-}^{} \lambda_{hG^+G^-}^{2} C_0[G^\pm, G^\pm, G^\pm] 
      -2\lambda_{HH^+ H^-}^{} \lambda_{hH^+H^-}^{2} C_0[H^\pm, H^\pm, H^\pm] \notag\\[8pt]
    &  -2\lambda_{HG^+ G^-}^{} \lambda_{hG^+H^-}^{2} C_0[G^\pm, H^\pm, G^\pm] 
    -2\lambda_{HG^+H^-}^{}\lambda_{hG^+H^-}^{}\lambda_{hG^+G^-}^{} (C_0[H^\pm, G^\pm, G^\pm] +C_0[G^\pm, G^\pm, H^\pm] ) \notag\\[8pt]
   & - 2\lambda_{HH^+H^-}^{}\lambda_{hG^+H^-}^2 C_0[H^\pm, G^\pm, H^\pm]
       -2\lambda_{HG^+H^-}^{}\lambda_{hH^+H^-}^{}\lambda_{hG^+H^-}^{}(C_0[H^\pm, H^\pm, G^\pm] + C_0[G^\pm, H^\pm, H^\pm]) \notag\\[8pt]
   & + 24\lambda_{Hhh}^{}\lambda_{hhhh}^{} B_0[q^2;h,h] 
      + 18\lambda_{hhh}^{}\lambda_{Hhhh}^{}(B_0[p_1^2; h, h] + B_0^{}[p_2^2; h,h] ) \notag\\[8pt]
   & + 12\lambda_{HHh}^{}\lambda_{Hhhh}^{}B_0[q^2; H, h] 
      + 8\lambda_{Hhh}^{}\lambda_{HHhh}^{} (B_0[p_1^2; H, h] + B_0[p_2^2; H, h]) \notag\\[8pt]
   & + 12 \lambda_{HHH}^{}\lambda_{HHhh}^{}B_0[q^2; H, H] 
     + 6\lambda_{HHh}^{}\lambda_{HHHh}^{}(B_0[p_1^2; H, H] + B_0^{}[p_2^2; H, H] ) \notag\\[8pt]
  & + 4\lambda_{HAA}^{}\lambda_{hhAA}^{} B_0[q^2; A, A] 
  + 2\lambda_{hAA}^{}\lambda_{HhAA}^{}(B_0[p_1^2; A,A]+ B_0[p_2^2; A, A])\notag\\[8pt]
  & + 4 \lambda_{HG^0 G^0}^{}\lambda_{hhG^0G^0}^{}B_0[q^2; G^0, G^0] 
     + 2\lambda_{hG^0G^0}^{}\lambda_{HhG^0G^0}^{}(B_0[p_1^2; G^0, G^0] + B_0[p_2^2; G^0, G^0] )\notag\\[8pt]
  & + 2\lambda_{HG^0A}^{}\lambda_{hhG^0A}^{} B_0[q^2; G^0, A] 
   + \lambda_{hG^0A}^{}\lambda_{HhG^0A}^{} ( B_0[p_1^2; G^0, A] + B_0[p_2^2; G^0, A] ) \notag\\[8pt]
  & + 2\lambda_{HH^+H^-}^{}\lambda_{hhH^+H^-}^{} B_0[q^2; H^\pm, H^\pm] 
    + \lambda_{hH^+H^-}^{}\lambda_{HhH^+H^-}^{}(B_0[p_1^2; H^\pm, H^\pm] + B_0[p_2^2; H^\pm, H^\pm])\notag\\[8pt]
  & + 2\lambda_{HG^+G^-}^{}\lambda_{hhG^+G^-}^{} B_0[q^2; G^\pm, G^\pm] 
    + \lambda_{hG^+G^-}^{}\lambda_{HhG^+G^-}^{}(B_0[p_1^2; G^\pm, G^\pm] + B_0[p_2^2; G^\pm, G^\pm])\notag\\[8pt]
  & + 4\lambda_{HG^+H^-}^{}\lambda_{hhG^+H^-}^{} B_0[q^2; H^\pm, G^\pm] 
    + 2\lambda_{hG^+H^-}^{}\lambda_{HhG^+H^-}^{}(B_0[p_1^2; H^\pm, G^\pm] + B_0[p_2^2; H^\pm, G^\pm])\notag\\[8pt]
  & + g^3 m_W^{}c_{\beta- \alpha}^{}(2B_0[q^2; W, W] - 1 )
      + \frac{g_Z^3 m_Z^{}}{2}c_{\beta- \alpha}^{}(2B_0[q^2; Z, Z] - 1 ), 
 \end{align}
 where definitions of combinations of $C^{}$-functions are given as
\begin{align}
  C_{SVV}^{\phi\phi\phi}[S, V_1, V_2]&= [p_1^2 C_{21}^{} + p_2^2C_{22}+2p_1\cdot p_2 C_{23} +4C_{24} -\frac{1}{2} \notag\\[8pt]
 &- (q+ p_1^{}) \cdot(p_1 C_{11} + p_2 C_{12}) + q\cdot p_1 C_0](S, V_1, V_2), 
 \end{align}
 \begin{align}
  C_{VSV}^{\phi\phi\phi}[V_2,S, V_1]&= [p_1^2 C_{21}^{} + p_2^2C_{22}+2p_1\cdot p_2 C_{23} +4C_{24} -\frac{1}{2} \notag\\[8pt]
 &+ (3p_1^{} -p_2) \cdot(p_1 C_{11} + p_2 C_{12}) + 2p_1\cdot (p_1 - p_2) C_0](V_2, S, V_1), \\[8pt]
  C_{VVS }^{\phi\phi\phi}[V_1, V_2, S]&= [p_1^2 C_{21}^{} + p_2^2C_{22}+2p_1\cdot p_2 C_{23} +4C_{24} -\frac{1}{2} \notag\\[8pt]
 &+ (3p_1^{} +4p_2) \cdot(p_1 C_{11} + p_2 C_{12}) + 2q\cdot (q +p_2) C_0](V_1, V_2, S), \\[8pt]
  C_{VSS}^{\phi\phi\phi}[V, S_1, S_2]&= [p_1^2 C_{21}^{} + p_2^2C_{22}+2p_1\cdot p_2 C_{23} +4C_{24} -\frac{1}{2} \notag\\[8pt]
  &+ (4p_1^{} +2p_2) \cdot(p_1 C_{11} + p_2 C_{12}) + 4q\cdot p_1  C_0](V, S_1, S_2), \\[8pt]
   C_{SVS}^{\phi\phi\phi}[S_2, V, S_1]& = [p_1^2 C_{21}^{} + p_2^2C_{22}+2p_1\cdot p_2 C_{23} +4C_{24} -\frac{1}{2} \notag\\[8pt]
  &+ 2p_2\cdot(p_1 C_{11} + p_2 C_{12}) -p_1\cdot (p_1 +2 p_2) C_0](S_2, V,S_1), \\[8pt]
  C_{SSV}^{\phi\phi\phi}[S_1, S_2, V]&= [p_1^2 C_{21}^{} + p_2^2C_{22}+2p_1\cdot p_2 C_{23} +4C_{24} -\frac{1}{2} \notag\\[8pt]
   &- 2p_2 \cdot(p_1 C_{11} + p_2 C_{12}) -q \cdot (p_1 - p_2) C_0](S_1, S_2, V),  
 \end{align}
 with
 \begin{align}
   C_i[X, Y, Z] \equiv C_i[p_1^2,p_2^2,q^2; m_X, m_Y, m_Z]. 
   \end{align}

\bibliographystyle{JHEP}
\bibliography{references}

\end{document}